\documentclass[aps,pre,reprint,superscriptaddress]{revtex4-1}

\usepackage[dvipdfmx]{graphicx}
\usepackage{amsmath}
\usepackage{dcolumn}
\usepackage{bm}
\usepackage{accents}
\usepackage{amssymb}
\usepackage{mathtools}
\usepackage{feynmp}
\usepackage{url}
\usepackage{setspace}
\usepackage{color}
\usepackage{pifont}
\usepackage{hyperref}

\usepackage{physics}

\newcommand{\revision}[1]{{{#1}}}

\begin{document}

\title{Thermodynamic uncertainty relation for feedback cooling}

\author{Kousuke Kumasaki}
\affiliation{Department of Applied Physics, The University of Tokyo, 7-3-1 Hongo, Bunkyo-ku, Tokyo 113-8656, Japan}

\author{Kaito Tojo}
\affiliation{Department of Applied Physics, The University of Tokyo, 7-3-1 Hongo, Bunkyo-ku, Tokyo 113-8656, Japan}

\author{Takahiro Sagawa}
\affiliation{Department of Applied Physics, The University of Tokyo, 7-3-1 Hongo, Bunkyo-ku, Tokyo 113-8656, Japan}
\affiliation{Quantum-Phase Electronics Center (QPEC), The University of Tokyo, 7-3-1 Hongo, Bunkyo-ku, Tokyo 113-8656, Japan}

\author{Ken Funo}
\affiliation{Department of Applied Physics, The University of Tokyo, 7-3-1 Hongo, Bunkyo-ku, Tokyo 113-8656, Japan}
\email{funo@ap.t.u-tokyo.ac.jp}

\date{\today}
\begin{abstract}
Feedback cooling enables a system to achieve low temperatures through measurement-based control. Determining the thermodynamic cost required to achieve the ideal cooling efficiency within a finite time remains an important problem. In this work, we establish a thermodynamic uncertainty relation (TUR) for feedback cooling in classical underdamped Langevin systems, thereby deriving a trade-off between the cooling efficiency and the entropy reduction rate. The obtained TUR implies that simultaneous achievement of the ideal cooling efficiency and finite entropy reduction rate is asymptotically possible by letting the fluctuation of the reversible local mean velocity diverge. This is shown to be feasible by using a feedback control based on the Kalman filter.
Our results clarify the thermodynamic costs of achieving the fundamental cooling limit of feedback control from the perspective of the TUR.
\end{abstract}

\maketitle

\section{Introduction} 
Recent experimental advances have enabled a precise investigation of thermodynamics in microscopic systems, where thermal fluctuations play a crucial role.
These systems require a theoretical framework beyond conventional thermodynamics, giving rise to stochastic thermodynamics~\cite{sekimoto2010stochastic, RevModPhys.81.1665, RevModPhys.83.771, Seifert_2012, PhysRevX.7.021051, Funo2018, RevModPhys.93.035008}, which explicitly incorporates thermal fluctuations.
In this framework, the second law constrains the \revision{average} entropy production to be nonnegative.
This framework has been further extended to information processing systems, known as information thermodynamics~\cite{Parrondo_Horowitz_Sagawa_2015, PhysRevLett.100.080403, PhysRevX.4.031015, PhysRevLett.128.170601, PhysRevLett.122.150603, yada2024experimentally, prech2025quantum}.
Although the second law represents a fundamental thermodynamic bound on physical processes, it does not give a tight bound on the entropy production in general finite-time processes.
To this end, thermodynamic uncertainty relations (TURs)~\cite{Barato_and_Seifert_2015_PhysRevLett.114.158101,
Gingrich_2016_PhysRevLett.116.120601, Horowitz_and_Gingrich_2017_PhysRevE.96.020103,
Dechant_2018,
Dechant_2018_2,
Vu_and_Hasegawa_2019_PhysRevE.100.032130,
Lee_2019_PhysRevE.100.062132} have been established as generalizations of the second law, providing bounds between entropy production and generalized currents, and revealing trade-offs between efficiency and power in thermal engines~\cite{Shiraishi_2016_PhysRevLett.117.190601, Pietzonka_and_Seifert_2018_PhysRevLett.120.190602, PhysRevLett.127.190604}.
TURs have also been extended to the setup of Maxwell's demon~\cite{Potts_2019, Van_Vu_2020, Liu_2020, Otsubo_PhysRevE.101.062106, Tanogami_PhysRevResearch.5.043280}, leading to the derivation of TURs involving partial entropy production~\cite{Otsubo_PhysRevE.101.062106, Tanogami_PhysRevResearch.5.043280}, and the establishment of trade-off relations in information heat engines~\cite{Tanogami_PhysRevResearch.5.043280}.

Feedback cooling is a paradigmatic example of an information heat engine, where work is extracted by converting information into energy.
Applying feedback based on measurement results enables the suppression of particle fluctuations. This technique has been realized in various experimental systems, including  nanomechanical resonators~\cite{Rossi_2018} and optically trapped nanoparticles~\cite{Ginseler_2012_PhysRevLett.109.103603, Magrini_2021, Gonzalez_Ballestero_2021, Kamba_shimizu_aikawa_2023, kamba2025quantum}.
While the standard formulation of the second law relates entropy production to the temperature of the heat bath,
the primary objective in feedback cooling is to reduce the temperature of the system below the bath temperature. 
In the study of classical systems, second-law-like inequalities have been established for Langevin dynamics~\cite{Horowitz_and_Sandberg_2014, Sandberg_2014_PhysRevE.90.042119}, relating the kinetic temperature of the system to the rate of information acquisition via measurement, quantified by the transfer entropy~\cite{PhysRevLett.85.461, Horowitz_and_Sandberg_2014, Hartich_2014}.
In feedback cooling, the effective use of information is crucial for achieving both efficient cooling and high cooling rates. Although examples achieving the fundamental information thermodynamic bound of feedback cooling have been reported~\cite{Horowitz_and_Sandberg_2014, kumasaki2025}, there exists a trade-off between power and efficiency, in general. Therefore, it is also expected that to achieve the ideal efficiency, the cooling power should vanish. 
However, without a TUR specific to feedback cooling, it has remained unclear whether maximum efficiency can coexist with a finite cooling rate, and, if so, what additional cost would be required.

\begin{figure}[btp]
\begin{center}
\includegraphics[width=0.7\linewidth]{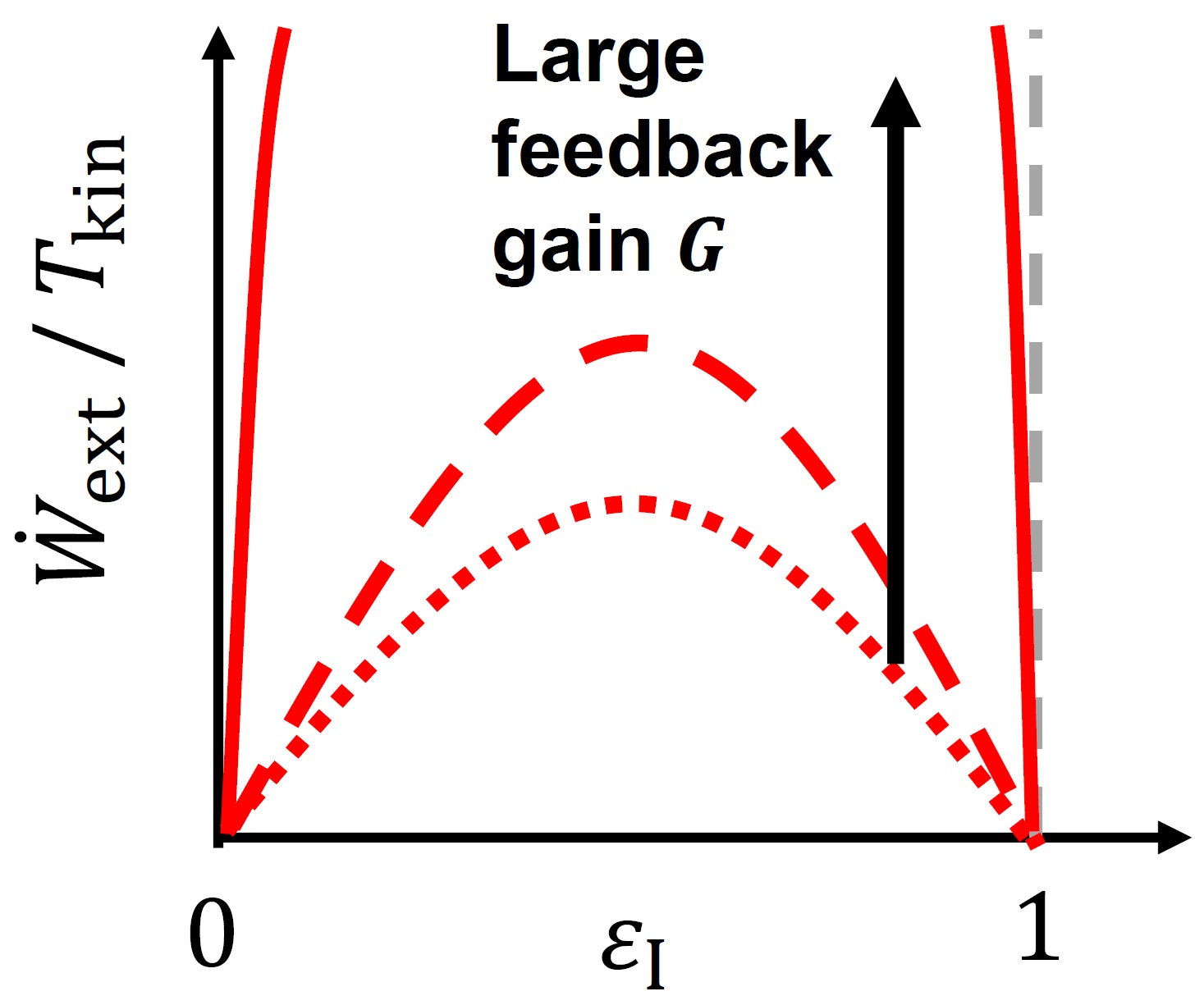}
\caption{Schematic of the TUR for feedback cooling, where $\epsilon_{\rm I}$ is the cooling efficiency and $\dot{W}_{\mathrm{ext}}/T_{\mathrm{kin}}$ is the entropy reduction rate. The region above the parabola is prohibited by the obtained TUR~\eqref{eq: tradeoff from TUR for feedback cooling}\revision{, where each parabola corresponds to the bound set by TUR for different parameter values (e.g., the feedback gain $G$)}. In the limit of large feedback gain $G\rightarrow \infty$, the entropy reduction rate maintains a finite value while approaching the maximum cooling efficiency $\epsilon_{\rm I}\rightarrow 1$ (see also Fig.~\ref{fig: achieving max efficiency}).   
}
\label{fig:TUR}
\end{center}
\end{figure}

In this \revision{paper}, we derive TUR for feedback cooling that describes a fundamental trade-off relation between the cooling efficiency and entropy reduction rate in underdamped bipartite Langevin systems.
The key technique to derive the main result is to introduce a novel orthogonal decomposition of the local mean velocity and the partial entropy production, in analogy with the decomposition proposed in~\cite{Dechant_PhysRevE.106.024125, Dechant_2022_2, Kamijima_2023_PhysRevE.107.L052101}, while our decomposition is unique to bipartite systems. 
This decomposition allows us to derive the TUR for feedback cooling by involving the kinetic temperature of the system, in contrast to the conventional TUR that involves the bath temperature. 
From the obtained TUR, we identify that divergence of the fluctuation of the reversible local mean velocity is required to simultaneously achieve the maximum cooling efficiency and finite entropy reduction rate. We introduce a concrete example based on the Kalman filter and show that increasing the feedback gain to infinity allows satisfying the above condition (see also Fig.~\ref{fig:TUR}). 
The obtained results provide design principles for achieving optimal feedback cooling, which is particularly relevant to nanoparticle experiments~\cite{Ginseler_2012_PhysRevLett.109.103603, Magrini_2021, Gonzalez_Ballestero_2021, Kamba_shimizu_aikawa_2023, kamba2025quantum}.  

\revision{This paper is organized as follows. In Sec.~\ref{sec: Setup}, we  introduce the setup of feedback cooling for a free particle.  
In Sec.~\ref{Sec: Orthogonal decomposition}, we present a new orthogonal decomposition of the mean local velocity and the partial entropy production rate.  In Sec.~\ref{sec: TUR}, we derive the main results, TURs for feedback cooling and a trade-off relation between the cooling efficiency and entropy reduction rate.
In Sec.~\ref{sec: Kalman example}, we use the Kalman filter and analyze the attainability of the ideal efficiency and finite cooling power based on the obtained TUR. 
In Sec.~\ref{sec: harmonic}, we discuss how the obtained results are generalized when we include the effect of the harmonic potential. We summarize our results in Sec.~\ref{sec: summary}.
}

\section{Setup}\label{sec: Setup}
\revision{We analyze the setup of feedback cooling by considering a bipartite system that consists of a system described by a Brownian particle and a memory that registers measurement outcomes~\cite{Horowitz_and_Sandberg_2014, Munakata_2013}.
The system is coupled to a thermal reservoir at temperature $T$ and is subject to a damping force $-\gamma v_{t}$, where $v_{t}$ is the velocity of the system.
A feedback force $f_{t}=-a y_{t}$ is applied to reduce the kinetic energy of the particle, where $y_{t}=(1/\tau)\int^{t}_{0} e^{-(t-s)/\tau}(v_{s}\dd s+\sigma \dd W^{y}_{s})$
denotes the outcome of continous measurement of the velocity by applying a low-pass filter with time constant $\tau$. 
Here, $\sigma^{2}$ represents the noise strength and $dW_{t}^{y}$ is a Wiener increment, which is a Gaussian random variable with zero mean and variance $dt$.
By following Refs.~\cite{Horowitz_and_Sandberg_2014, Munakata_2013}, the corresponding Langevin equation reads
\begin{align} \label{eq: LE sys}
    m \dd v_{t} &= -\gamma v_{t}\dd t -a y_{t}\dd t + \sqrt{2\gamma T} \dd W_{t} , \\
    \tau \dd y_{t} &= -(y_{t}\dd t-v_{t}\dd t-\sigma \dd W_{t}^{y}),
\end{align}
where $\dd W_{t}$ is another Wiener increment which is independent from $\dd W_{t}^{y}$. 
The corresponding Fokker-Planck equation for the joint probability distribution $p(v, y)$ reads:}
\begin{align}\label{eq: FP eq for HS}
\partial_t p(v,y) = -\partial_v J_v(v,y) - \partial_y J_y(v,y).
\end{align}
The $v$ and $y$ components of the probability current are given by
\begin{align}
J_{v}(v,y) &= \underbrace{-\frac{ay}{m} p(v,y)}_{J_v^{\mathrm{rev}}(v,y)} \ \underbrace{-\frac{\gamma}{m} \Bigl( v+\frac{T}{m}\partial_{v}\Bigr) p(v,y)}_{J_v^{\mathrm{irr}}(v,y)}   , \label{eq: FP eq for v} \\
J_y(v,y) &= -\frac{1}{\tau}(y - v) p(v,y) - \frac{\sigma^2}{2\tau^2} \partial_y p(v,y), \label{eq: FP eq for y}
\end{align}
where we set the Boltzmann constant $k_{\mathrm{B}}$ to unity.
In Eq.~\eqref{eq: FP eq for v}, we further decompose the probability current $J_v(v, y)$ into reversible $J_v^{\mathrm{rev}}(v,y)$ and irreversible $J_v^{\mathrm{irr}}(v,y)$ contributions~\cite{Risken1996, Spinney2012} as $J_v(v, y)=J_v^{\mathrm{rev}}(v,y)+J_v^{\mathrm{irr}}(v,y)$, \revision{where the feedback force $-ay$ arising from an external potential is treated as a reversible contribution (see Ref.~\cite{Horowitz_and_Sandberg_2014, Munakata_2013} for detailed discussions).
Note that Eq.~\eqref{eq: FP eq for v} describes the case of a free particle, but our results can be generalized to the case where the particle is trapped in a harmonic potential, as discussed in Sec.~\ref{sec: harmonic}.}

We now introduce thermodynamic quantities based on the dynamics described by Eq.~\eqref{eq: FP eq for HS}.
In the context of feedback cooling, a key quantity is the kinetic energy of the system defined as
\revision{
\begin{align}
    E \coloneqq 
      \int \frac{1}{2}m v^2 p(v, y)\, \dd v \dd y.
\end{align}
}
From the first law of thermodynamics $
\dot{E} = -\dot{W}_{\mathrm{ext}} + \dot{Q}
$, we define the heat current and the extractable work rate using the irreversible and reversible currents as
\revision{~\cite{sekimoto2010stochastic, Rosinberg2015}
\begin{align}
    \dot{Q} & \coloneqq \int m v J_v^{\mathrm{irr}}(v,y)\, \dd v\dd y, \\
    \dot{W}_{\mathrm{ext}} & \coloneqq -\int m v J_v^{\mathrm{rev}}(v,y)\, \dd v\dd y.    
\end{align}
}The generalized second law of thermodynamics is given by\revision{~\cite{PhysRevX.4.031015, Rosinberg_2016}
\begin{align}
\dot{\Sigma}_{\nu^{\rm irr}} \coloneqq \dot{s}_v - \dot{Q}/T - \dot{I}_v  \geq 0, \label{eq: SL}
\end{align}
}where $\dot{\Sigma}_{\nu^{\rm irr}}$ is the partial entropy production rate of the system, $\dot{s}_v \coloneqq -\int J_v(v) \partial_v \ln \tilde{p}(v) \, \dd v$ is the Shannon entropy rate of the system, $\tilde{p}(v) \coloneqq \int p(v,y)\, \dd y$ is the marginal probability distribution of the system, and
\begin{align}
    \dot{I}_v \coloneqq -\int (\partial_v J_v(v,y)) \ln \frac{p(v,y)}{\tilde{p}(v)}\, \dd v\dd y
\end{align}
is the information flow, quantifying the change in the system-memory correlation induced by the time evolution of the system.
\revision{
By using the irreversible local mean velocity $\nu_v^{\mathrm{irr}}(v,y) \coloneqq J_v^{\mathrm{irr}}(v,y)/p(v,y)$, we can express the partial entropy production rate as the square of the norm of $\nu_v^{\mathrm{irr}}$ as~\cite{PhysRevX.4.031015, Rosinberg_2016}
\begin{align}
\dot{\Sigma}_{\nu^{\rm irr}} = \frac{m^2}{\gamma T}\int (\nu_v^{\mathrm{irr}}(v, y))^2 p(v, y) = \| \nu_v^{\mathrm{irr}} \|_p^{2} \geq 0, \label{eq: SL2}
\end{align}
which directly shows the nonnegativity of $\dot{\Sigma}_{\nu^{\rm irr}}$. Here, we introduce the norm $\|\cdot\|_p\coloneqq \langle\cdot,\cdot\rangle_{p}^{1/2}$ associated with the inner product for functions $f(v,y)$ and $g(v,y)$, defined as}
\revision{
\begin{equation}
    \langle f, g \rangle_p \coloneqq \frac{m^2}{\gamma T} \int f(v,y) g(v,y) p(v,y)\, \dd v\dd y.
    \label{eq: inner_product}
\end{equation}
}

\begin{figure}[tbp]
\begin{center}
\includegraphics[width=0.95\linewidth]{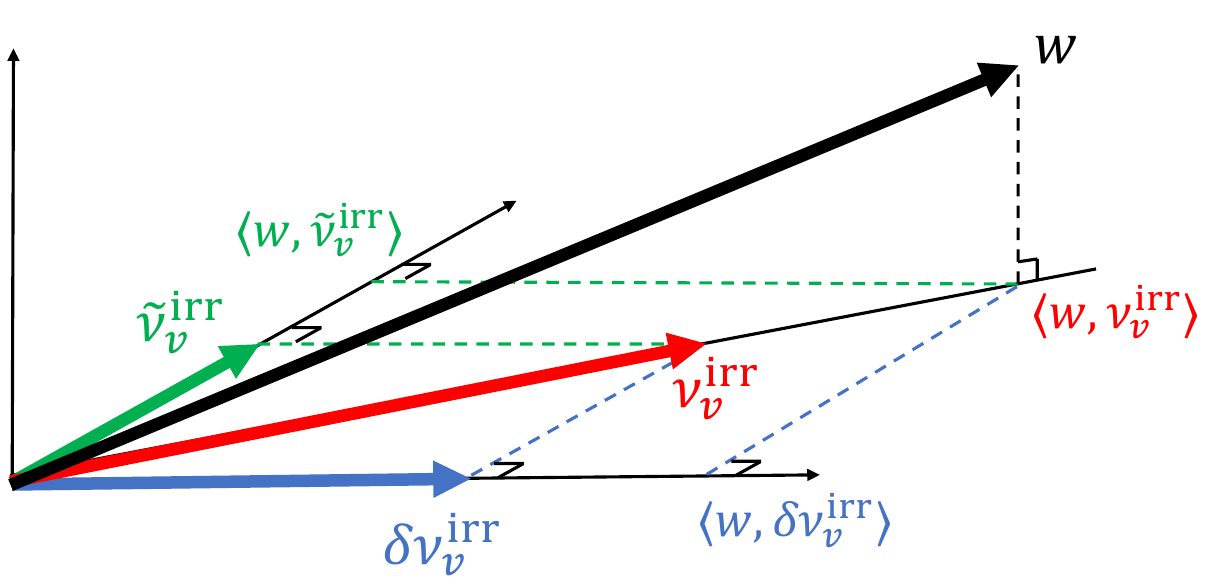}
\caption{Schematic of the orthogonal decomposition used in this study.  
The irreversible local mean velocity $\nu_v^{\mathrm{irr}}$ is decomposed into two orthogonal components, which are  
a marginalized part $\tilde{\nu}_v^{\mathrm{irr}}$ and a residual part $\delta \nu_v^{\mathrm{irr}}$.  
The irreversible component of a generalized current with a general observable $w(v,y)$ is given by  
$\mathcal{J}^{w}_{\nu^{\rm irr}} \coloneqq (\gamma T/m^2) \ev{w, \nu_v^{\mathrm{irr}}}_p$.  
Using the decomposition  
$\ev{w, \nu_v^{\mathrm{irr}}}_p = \ev{w, \delta \nu_v^{\mathrm{irr}}}_p + \ev{w, \tilde{\nu}_v^{\mathrm{irr}}}_p$,  
the current $\mathcal{J}^{w}_{\nu^{\rm irr}}$ is decomposed into each component. Each term leads to distinct TURs~\eqref{eq: short time TUR for partial entropy production}-\eqref{eq: short time TUR for marginal entropy production}. 
}
\label{fig:orthogonal_decomposition}
\end{center}
\end{figure}

\section{\label{Sec: Orthogonal decomposition}Orthogonal decomposition of the partial entropy production}
In this section, we introduce a new orthogonal decomposition of the partial entropy production rate $\dot{\Sigma}_{\nu^{\rm irr}}$ 
\revision{with respect to the inner product~\eqref{eq: inner_product}}
and show that one of its components reproduces the second-law-like inequality in the previous study~\cite{Horowitz_and_Sandberg_2014}. 

We begin by decomposing the irreversible local mean velocity into orthogonal components as
\begin{align}
    \nu_v^{\mathrm{irr}} = \delta \nu_v^{\mathrm{irr}} + \tilde{\nu}_v^{\mathrm{irr}}, \label{eq: velocity decomposition}
\end{align}
where 
\revision{
\begin{eqnarray}
    \tilde{\nu}_v^{\mathrm{irr}}(v) &\coloneqq& \int J_v^{\mathrm{irr}}(v,y)/\tilde{p}(v) \, \dd y \nonumber\\
&=& - \frac{\gamma}{m} \left(v + \frac{T}{m} \partial_v \ln \tilde{p}(v) \right)
\end{eqnarray}
is the marginalized irreversible local mean velocity. }
The residual component $\delta \nu_v^{\mathrm{irr}}$ is defined by Eq.~\eqref{eq: velocity decomposition}, which reads
\revision{
\begin{align}
    \delta \tilde{\nu}_v^{\mathrm{irr}}(v, y) \coloneqq \nu_v^{\mathrm{irr}}(v, y) - \tilde{\nu}_v^{\mathrm{irr}}(v) 
= - \frac{\gamma T}{m^2} \partial_v \ln \frac{p(v, y)}{\tilde{p}(v)}.
\end{align}
} 
From direct calculation, $\tilde{\nu}_v^{\mathrm{irr}}$ and $\delta \nu_v^{\mathrm{irr}}$ satisfy the orthogonality condition $
\langle \delta \nu_v^{\mathrm{irr}}, \tilde{\nu}_v^{\mathrm{irr}} \rangle_p = 0$, meaning that the decomposition~\eqref{eq: velocity decomposition} is indeed orthogonal  (see Fig.~\ref{fig:orthogonal_decomposition}). Note that this decomposition is distinct from the Hatano-Sasa~\cite{Hatano_and_Sasa_2001_PhysRevLett.86.3463} and Maes-Neto\v{c}n\'{y}~\cite{Maes_and_Netocny_2014} decompositions and is unique to bipartite systems.

Based on the decomposition~\eqref{eq: velocity decomposition}, the partial entropy production rate can be decomposed as
\begin{align}\label{eq: EP decomp}
\dot{\Sigma}_{\nu^{\rm irr}} = \dot{\Sigma}_{\delta \nu^{\rm irr}} + \dot{\Sigma}_{\tilde{\nu}^{\rm irr}},
\end{align}
\revision{where $\dot{\Sigma}_{\delta \nu^{\rm irr}} \coloneqq \| \delta \nu_v^{\mathrm{irr}} \|_p^2$ and $\dot{\Sigma}_{\tilde{\nu}^{\rm irr}} \coloneqq \| \tilde{\nu}_v^{\mathrm{irr}} \|_p^2$. 

We now discuss the physical meaning of the decomposition Eq.~\eqref{eq: EP decomp} by assuming the typical setup of Gaussian feedback cooling. Specifically, we assume that the initial distribution $p_{0}(v, y)$ is a Gaussian distribution centered at the origin, i.e., its first moments vanish.
Note that for such initial distribution, the joint probability distribution $p(v, y)$ at later times remains Gaussian centered at the origin under the dynamics described by Eqs.~\eqref{eq: FP eq for HS},~\eqref{eq: FP eq for v} and \eqref{eq: FP eq for y}. 
Then, we can relate the decomposed partial entropy production rates with the kinetic temperature of the system $T_{\mathrm{kin}} \coloneqq m \ev{v^2}$ as}
\begin{align}
\dot{\Sigma}_{\delta \nu^{\rm irr}} &= \| \delta \nu_v^{\mathrm{irr}} \|_p^2 = \dot{s}_v - \frac{\dot{Q}}{T_{\mathrm{kin}}} - \dot{I}_v\geq 0, \label{eq: definition of delta S}\\
\dot{\Sigma}_{\tilde{\nu}^{\rm irr}} &= \| \tilde{\nu}_v^{\mathrm{irr}} \|_p^2 = \frac{\dot{Q}}{T_{\mathrm{kin}}} - \frac{\dot{Q}}{T}\geq 0. \label{eq: definition of tilde S}
\end{align}
\revision{To derive Eq.~\eqref{eq: definition of tilde S}, we use the Gaussian assumption to describe the marginal probability distribution in terms of the kinetic tempearature as $\tilde{p}(v) \propto \exp[-mv^2/(2T_{\mathrm{kin}})]$. Then, we have $\tilde{\nu}_{v}^{\rm irr}=(\gamma v/m)(T/T_{\mathrm{kin}} - 1)$ and $\| \tilde{\nu}_v^{\mathrm{irr}} \|_p^2= (\gamma/m) (T-T_{\rm kin})(T_{\rm kin}^{-1} -T^{-1})$. On the other hand, the heat current reads $\dot{Q}=\int mv\tilde{\nu}_{v}^{\rm irr}(v)\tilde{p}(v) \dd v = (\gamma/m) (T-T_{\rm kin})$, which completes the proof of Eq.~\eqref{eq: definition of tilde S}. Note that Eq.~\eqref{eq: definition of delta S} directly follows from the expression $\| \delta \nu_v^{\mathrm{irr}} \|_p^2 = \| \nu_v^{\mathrm{irr}} \|_p^2 - \| \tilde{\nu}_v^{\mathrm{irr}} \|_p^2$ combined with Eqs.~\eqref{eq: SL}, \eqref{eq: SL2} and \eqref{eq: definition of tilde S}.}

Since $\dot{\Sigma}_{\nu^{\rm irr}} \geq \dot{\Sigma}_{\delta \nu^{\rm irr}}$ and $\dot{\Sigma}_{\nu^{\rm irr}} \geq \dot{\Sigma}_{\tilde{\nu}^{\rm irr}} $, inequalities~\eqref{eq: definition of delta S} and \eqref{eq: definition of tilde S} give tighter bounds compared to the generalized second law~\eqref{eq: SL}. In addition, inequality~\eqref{eq: definition of tilde S} shows that achieving $T_{\mathrm{kin}} < T$ requires $\dot{Q} \geq 0$, which means that cooling requires heat dissipation within this setup. On the other hand, inequality~\eqref{eq: definition of delta S} is similar to~\eqref{eq: SL}, but the bath temperature $T$ is replaced by the kinetic temperature $T_{\rm kin}$, which allows quantifying the fundamental cooling limit from the perspective of information thermodynamics as follows. 
Using the relation $\dot{s}_v= \dot{E}/T_{\rm kin}= (\dot{Q}-\dot{W}_{\mathrm{ext}})/T_{\mathrm{kin}}$, inequality~\eqref{eq: definition of delta S} is equivalent to
\begin{align}\label{eq: information flow cooling limit}
\frac{\dot{W}_{\mathrm{ext}}}{T_{\mathrm{kin}}} \leq -\dot{I}_v,
\end{align}
\revision{where the left-hand side quantifies the entropy extracted from the system via feedback, which we refer to as the entropy reduction rate.
This can be regarded as an indicator of how effectively the system is cooled by extracting work and reducing its kinetic temperature.}
Therefore, inequality~\eqref{eq: information flow cooling limit} characterizes the fundamental limit of \revision{Gaussian} feedback cooling by setting an upper bound on the entropy reduction rate in terms of the information flow.
Assuming that the energy of the system is reduced by feedback, that is, when the cooling condition $\dot{W}_{\mathrm{ext}} \geq 0$ is satisfied, we define the information-to-entropy reduction cooling efficiency as\revision{
\begin{align}
    \varepsilon_{\mathrm{I}} \coloneqq \left| \frac{\dot{W}_{\mathrm{ext}}}{T_{\mathrm{kin}} \dot{I}_v} \right| \leq 1,
\end{align}
}where the last inequality follows from~\eqref{eq: information flow cooling limit}.

In the steady state, the information flow is always bounded by the \revision{transfer entropy rate~\cite{PhysRevLett.85.461, Hartich_2014} 
\begin{align}
\dot{I}_{\mathrm{TE}}\coloneqq\lim_{\dd t\rightarrow 0}\frac{1}{\dd t}\mathbb{E}_{Y_{t}}[s_{v}(p_{v}^{Y_{t}})-\mathbb{E}_{y_{t + \dd t}}[s_{v}(p_{v}^{Y_{t+\dd t}})]]
\end{align} via 
\begin{align}\label{eq: inequality information}
    -\dot{I}_v \leq \dot{I}_{\mathrm{TE}},
\end{align}
where $Y_{t}\coloneqq\{y_{s}\}_{[0,t)}$ and $Y_{t+\dd t}\coloneqq\{y_{s}\}_{[0,t]}$ are the collection of measurement results until $t$ and $t+\dd t$, respectively. We have further introduced the probability distribution of the system conditioned on the measurement outcomes $Y_{t}$ as $p_{v}^{Y_{t}}$, and $s_{v}$ is the Shannon entropy. From this expression, the transfer entropy rate quantifies the additional information obtained from the measurement result $y_{t}$ under the condition that $Y_{t}$ is already known. Because $\dot{I}_{\rm TE}$ quantifies the maximum information that can be utilized for feedback control, it gives an upper bound on the information flow~\eqref{eq: inequality information}, and a sufficient condition for the equality to hold is by using the Kalman filter~\cite{Horowitz_and_Sandberg_2014}.

Using the above relation between $\dot{I}_{v}$ and $\dot{I}_{\rm TE}$ in the steady state, inequality~\eqref{eq: information flow cooling limit} reproduces the second-law-like inequality for feedback cooling~\cite{Horowitz_and_Sandberg_2014}: 
\begin{align} \label{eq: information flow cooling limit2}
\frac{\dot{W}_{\mathrm{ext}} }{ T_{\mathrm{kin}} } \leq \dot{I}_{\mathrm{TE}},
\end{align}
where the equality is acheivable by using the Kalman filter and sufficiently large feedback gain. 
}


\section{\label{sec: TUR}TUR for feedback cooling}
For orthogonal decompositions of the entropy production, a short-time TUR can be derived for each decomposed component~\cite{Dechant_PhysRevE.106.024125}.  
In the same spirit, we derive TURs corresponding to $\dot{\Sigma}_{\nu^{\rm irr}}$, $\dot{\Sigma}_{\delta \nu^{\rm irr}}$, and $\dot{\Sigma}_{\tilde{\nu}^{\rm irr}}$ based on the new orthogonal decomposition as follows. 

We begin by defining the generalized current of the system as $\mathcal{J}^{w}_{\nu} \coloneqq (\gamma T/m^2) \ev{w, \nu_v}_p$, where $\nu_{v}\coloneqq J_{v}(v,y)/p(v,y)$ is the local mean velocity, and $w(v,y)$ is a general observable. Since only the irreversible part of the local mean velocity $\nu_v^{\mathrm{irr}}$ is related to the (partial) entropy production $\dot{\Sigma}_{\nu^{\rm irr}}$ for underdamped Langevin systems, we focus on the irreversible component of the generalized current, defined as $
\mathcal{J}^{w}_{\nu^{\rm irr}} \coloneqq (\gamma T/m^2) \ev{w, \nu_v^{\mathrm{irr}}}_p$. Using the orthogonal decomposition~\eqref{eq: velocity decomposition}, we obtain the following decomposition for the irreversible current (see also Fig.~\ref{fig:orthogonal_decomposition}):
\begin{align}
\mathcal{J}^{w}_{\nu^{\rm irr}} &= \mathcal{J}^{w}_{\delta\nu^{\rm irr}} + \mathcal{J}^{w}_{\tilde{\nu}^{\rm irr}}, \label{eq: decomposition of the ireversible generalized current}
\end{align}
where $\mathcal{J}^{w}_{\delta\nu^{\rm irr}}\coloneqq (\gamma T/m^2) \ev{w, \delta \nu^{\mathrm{irr}}_v}_p$ and $\mathcal{J}^{w}_{\tilde{\nu}^{\rm irr}}\coloneqq (\gamma T/m^2) \ev{w, \tilde{\nu}^{\mathrm{irr}}_v}_p$. 
Because each current that appears in Eq.~\eqref{eq: decomposition of the ireversible generalized current} is expressed in terms of inner products, 
we can apply the Cauchy-Schwarz inequality \revision{$|\ev{f, g}_p|^2 \leq \|f\|_p^2 \|g\|_p^2$} to derive short-time TURs~\cite{Dechant_PhysRevE.106.024125}.
This yields the following set of inequalities:
\begin{align}
\left(\mathcal{J}^{w}_{\nu^{\rm irr}}\right)^2 &\leq \revision{ \left( \frac{\gamma T}{m^2}\right)^{2} } \left\| w \right\|_p^2 \dot{\Sigma}_{\nu^{\rm irr}}, \label{eq: short time TUR for partial entropy production} \\
\left(\mathcal{J}^{w}_{\delta\nu^{\rm irr}}\right)^2 &\leq \revision{ \left( \frac{\gamma T}{m^2}\right)^{2} } \left\| w \right\|_p^2 \dot{\Sigma}_{\delta \nu^{\rm irr}}, \label{eq: short time TUR for feedback cooling} \\
\left(\mathcal{J}^{w}_{\tilde{\nu}^{\rm irr}}\right)^2 &\leq \revision{ \left( \frac{\gamma T}{m^2}\right)^{2} } \left\| w \right\|_p^2 \dot{\Sigma}_{\tilde{\nu}^{\rm irr}}. \label{eq: short time TUR for marginal entropy production}
\end{align}
Here, inequality~\eqref{eq: short time TUR for partial entropy production} is similar to bipartite TURs for overdamped Langevin systems~\cite{Otsubo_PhysRevE.101.062106, Tanogami_PhysRevResearch.5.043280}. However, unlike the overdamped case, the irreversible current $\mathcal{J}^{w}_{\nu^{\rm irr}}$ deviates from the total current $\mathcal{J}^{w}_{\nu}$ because of the presence of reversible contributions in underdamped systems.
Next, the partial entropy production rate $\dot{\Sigma}_{\delta \nu^{\rm irr}}$ that appears in inequality~\eqref{eq: short time TUR for feedback cooling} corresponds to the information thermodynamic cooling limit~\eqref{eq: definition of delta S} and~\eqref{eq: information flow cooling limit}. Therefore, it can be interpreted as a TUR associated with feedback cooling as we discuss in detail in the following.  
Finally, inequality~\eqref{eq: short time TUR for marginal entropy production} shows that $\mathcal{J}^{w}_{\tilde{\nu}^{\rm irr}}$ must vanish in the limit of $T_{\mathrm{kin}} \to T$ by using~\eqref{eq: definition of tilde S}.

\revision{Note that we do not require Gaussian assumptions for the derivation of the TURs~\eqref{eq: short time TUR for partial entropy production}-\eqref{eq: short time TUR for marginal entropy production}, and they can be readily generalized to the case with trapping potentials [see Eqs.~\eqref{eq: short time TUR for partial entropy production with harmonic potential}-\eqref{eq: short time TUR for marginal entropy production with harmonic potential}]. In addition, generalization to higher dimensions is straightforward.}

We now focus on the trade-off implied by~\eqref{eq: short time TUR for feedback cooling}.  
As bipartite TURs lead to a trade-off between power and efficiency in information engines~\cite{Tanogami_PhysRevResearch.5.043280}, we derive a trade-off relation between the entropy reduction rate and cooling efficiency based on~\eqref{eq: short time TUR for feedback cooling}.
To this end, we choose $w(v,y) = \nu_v(v,y) - \tilde{\nu}_v^{\mathrm{irr}}(v)$, which allows us to express the information flow as 
\begin{align}
\revision{ \mathcal{J}^{w}_{\delta \nu^{\rm irr}} } = \frac{\gamma T}{m^2}\ev{\nu_{v}-\tilde{\nu}_{v}^{\rm irr}, \revision{\delta \nu_v^{\mathrm{irr}}} }_p = -\dot{I}_v. \label{eq: choice of J}
\end{align}
Substituting Eq.~\eqref{eq: choice of J} into~\eqref{eq: short time TUR for feedback cooling}, and assuming the cooling condition $\dot{W}_{\rm ext}\geq 0$ \revision{and the initial distribution to be Gaussian centered at the origin}, we obtain the main result that gives a trade-off relation between the cooling efficiency and entropy reduction rate \revision{for Gaussian systems}:
\begin{align}
 \frac{\dot{W}_{\mathrm{ext}}}{T_{\mathrm{kin}}}  \leq \revision{ \left( \frac{\gamma T}{m^2} \right)^{2} } \left\| \nu_{v} - \tilde{\nu}_v^{\mathrm{irr}} \right\|_p^2 \varepsilon_{\mathrm{I}} (1 - \varepsilon_{\mathrm{I}}).\label{eq: tradeoff from TUR for feedback cooling}
\end{align}
Inequality~\eqref{eq: tradeoff from TUR for feedback cooling} indicates that achieving the maximum cooling efficiency $\varepsilon_{\mathrm{I}} \to 1$ with finite entropy reduction rate requires
$
\revision{(\gamma T/m^2)^2} \left\| \nu_v - \tilde{\nu}_v^{\mathrm{irr}} \right\|_p^2 \to \infty
$ (see also Fig.~\ref{fig:TUR}).
This quantity represents the fluctuation of the generalized current associated with $w= \nu_v - \tilde{\nu}_v^{\mathrm{irr}}$ and satisfies\revision{
\begin{align}\label{eq: variation of information flow in tilde}
  \left\| \nu_v - \tilde{\nu}_v^{\mathrm{irr}} \right\|_p^2 =  \| \nu_{v}^{\mathrm{rev}} \|_p^2 + 2 \langle \nu_{v}^{\mathrm{rev}}, \delta \nu_v^{\mathrm{irr}} \rangle_p + \| \delta \nu_v^{\mathrm{irr}} \|_p^2 ,
\end{align}
where $\nu_{v}^{\mathrm{rev}}(v,y)\coloneqq J_{v}^{\rm rev}(v,y)/p(v,y)$}.
Before discussing Eq.~\eqref{eq: variation of information flow in tilde}, we note that based on the bipartite TUR for overdamped systems, achieving finite power at maximum efficiency in information heat engines requires divergence of the short-time fluctuation of the information flow $\revision{ (\gamma T/m^2)^{2} }\| \delta \nu_v^{\mathrm{irr}} \|_p^2$~\cite{Tanogami_PhysRevResearch.5.043280}. On the other hand, in our feedback cooling setup, achieving the maximum cooling efficiency (i.e., the equality condition of~\eqref{eq: definition of delta S}) implies that $\| \delta \nu_v^{\mathrm{irr}} \|_p^2$ vanishes.
Nevertheless, if the reversible contribution $\| \nu_v^{\mathrm{rev}} \|_p^2$ or $\langle \nu_{v}^{\mathrm{rev}}, \delta \nu_v^{\mathrm{irr}} \rangle_p$ diverges, the obtained TUR allows the possibility of achieving maximum efficiency with finite entropy reduction rate.



\section{\label{sec: Kalman example}Achieving the maximum  cooling efficiency with finite entropy reduction rate}
\begin{figure}[tbp]
\centering
\includegraphics[width=\linewidth]{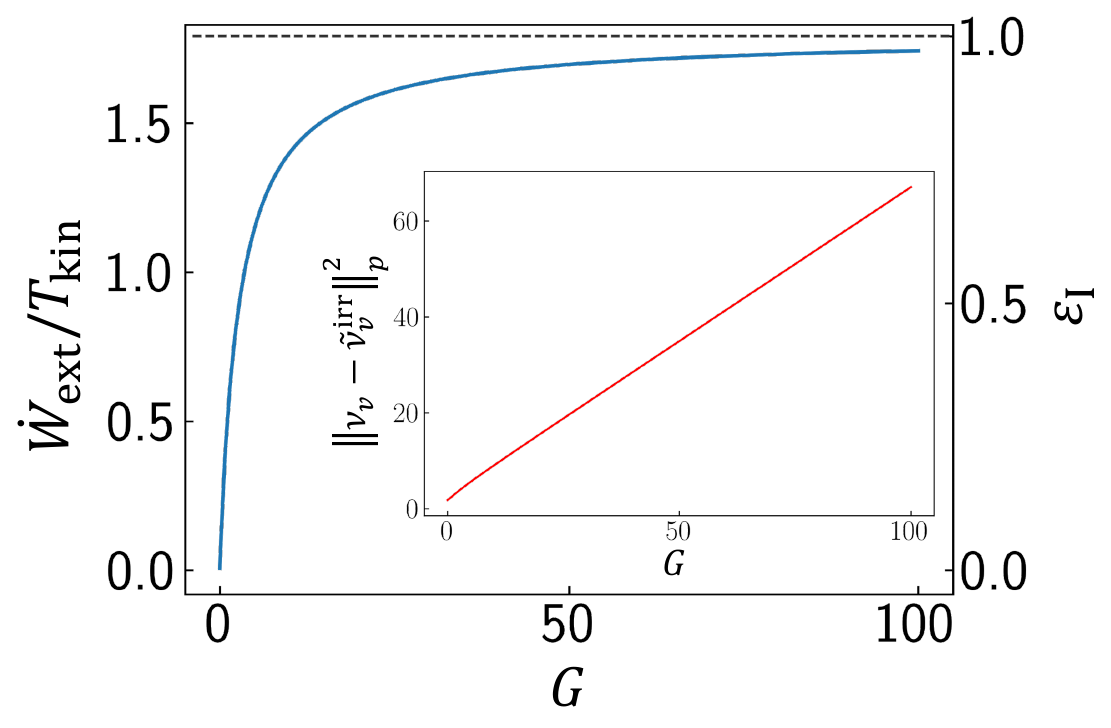}
\caption{Behavior of the entropy reduction rate and the cooling efficiency as functions of the feedback gain $G$.
The main plot shows the entropy reduction rate $\dot{W}_{\mathrm{ext}}/T_{\mathrm{kin}}$ (left axis) and the cooling efficiency $\varepsilon_{\mathrm{I}}$ (right axis)\revision{, where the two plots overlap with each other because the information flow $\dot{I}_{v}=-K/2m$ does not depend on $G$.}
As $G$ increases, $\varepsilon_{\mathrm{I}}$ asymptotically approaches unity, while the fluctuation associated with the entropy reduction rate (shown in the inset) diverges linearly.
This divergence clarifies how asymptotically achieving finite entropy reduction rate at maximum efficiency is possible from the viewpoint of the obtained TUR~\eqref{eq: tradeoff from TUR for feedback cooling}. The parameters are: $m=1$, $\gamma=1$, $T=5$, and $\sigma^2=0.5$.}
\label{fig: achieving max efficiency}
\end{figure}

We now demonstrate an explicit example in which the reversible contribution $\| \nu_v^{\mathrm{rev}} \|_p^2$ diverges, and the ideal cooling efficiency $\varepsilon_{\mathrm{I}} = 1$ is asymptotically achieved with finite entropy reduction rate. \revision{As suggested from the analysis based on the second-law-like inequality \eqref{eq: information flow cooling limit2}, the ideal cooling efficiency is achieved by using the Kalman filter with large feedback gain~\cite{Horowitz_and_Sandberg_2014}.}
This is realized in the dynamics governed by Eq.~\eqref{eq: FP eq for HS}, which can be reduced to a feedback control based on the Kalman filter~\cite{simon2006optimal} by introducing the rescaled variable
$
\hat{v} =  K/(\gamma + K + G) y,
$
and choosing the parameters as
$
a = G K/(\gamma + K + G)$ and $\tau = m/(\gamma + K + G)$. Here, $K$ and $G$ are the Kalman gain and the feedback gain, respectively. We further choose the so-called optimal Kalman gain by setting $K = \gamma ( \sqrt{1 + 2T/\sigma^2 \gamma} - 1 )$. 
Under these settings, $\hat{v}$ corresponds to the Kalman estimate of the particle velocity~\cite{Horowitz_and_Sandberg_2014}.
In the following, we consider the steady state. Then, the covariance matrix of the velocity $v$ and its estimate $\hat{v}$ is given by~\cite{Horowitz_and_Sandberg_2014} 
\begin{align}
\bm{\Xi} = \begin{pmatrix}
\sigma_{vv} & \sigma_{v \hat{v}} \\
\sigma_{v \hat{v}} & \sigma_{\hat{v} \hat{v}}
\end{pmatrix}
= \begin{pmatrix}
\mathcal{V}  + \mathcal{E} & \mathcal{V} \\
\mathcal{V} & \mathcal{V}
\end{pmatrix},
\end{align}
where
$\mathcal{V}= K\mathcal{E}/[2(\gamma+G)]$ and $ \mathcal{E} = K \sigma^2/m$. 
\revision{Using $\mathcal{V}$ and $\mathcal{E}$, the bath temperature and the kinetic temperature read $T=(m\mathcal{E}/2)(K/\gamma +2)$ and $T_{\rm kin}=m(\mathcal{V}+\mathcal{E})$, respectively. We also note that $\dot{W}_{\rm ext}=\dot{Q}=(\gamma/m)(T-T_{\rm kin})=G\mathcal{V}$. In addition, $\delta\nu_{v}^{\rm irr}=-(\gamma T/m^{2})(\hat{v}/\mathcal{E}-v\mathcal{V}/[\mathcal{E}(\mathcal{V}+\mathcal{E})] )$, and the information flow reads $\dot{I}_{v}=(\gamma/m)(1-T/(m\mathcal{E}))=-K/2m$. Using these expressions,}
the decomposed partial entropy production rate becomes
\begin{align}
\dot{\Sigma}_{\delta \nu^{\mathrm{irr}}}=\frac{\gamma T}{m^{2}}\Bigl(  \frac{1}{\mathcal{E}} - \frac{1}{\mathcal{V} + \mathcal{E}}\Bigr) \rightarrow 0 , \quad (G \to \infty),
\end{align}
and therefore, $\varepsilon_{\mathrm{I}} \rightarrow 1$. 
On the other hand, the entropy reduction rate asymptotically converges to a finite value:
\begin{align}\label{eq: Kalman entropy reduction rate}
\frac{\dot{W}_{\mathrm{ext}}}{T_{\mathrm{kin}}} = \frac{G \mathcal{V}}{m (\mathcal{V}+\mathcal{E})} \rightarrow \frac{K}{2m}, \quad (G \to \infty).
\end{align}
Thus, in the limit of large feedback gain $G$, the ideal efficiency is achieved with finite entropy reduction rate.
According to inequality~\eqref{eq: tradeoff from TUR for feedback cooling}, this implies that the fluctuation term $ \| \nu_v - \tilde{\nu}_v^{\mathrm{irr}} \|_p^2$ must diverge.  
This can be explicitly verified as:\revision{
\begin{align}
  \left\| \nu_v -  \tilde{\nu}_v^{\mathrm{irr}} \right\|_p^2 & =  \left\| \nu_v^{\rm rev} + \delta \nu_v^{\mathrm{irr}} \right\|_p^2  \notag \\
=&   \frac{G^2 \mathcal{V}}{\gamma T}  + \frac{2G}{m} \frac{\mathcal{V} }{ \mathcal{V} +\mathcal{E}} +    \dot{\Sigma}_{\delta\nu^{\rm irr}} \notag \\
\rightarrow &  \infty \quad (G \to \infty),
\end{align} 
where the divergent term comes from the reversible contribution: $\| \nu_v^{\mathrm{rev}} \|_p^2= G^{2}\mathcal{V}/(\gamma T) =O(G)$.} 
In Fig.~\ref{fig: achieving max efficiency}, we plot the asymptotic achievement of the ideal efficiency $\varepsilon_{\mathrm{I}} = 1$.  
The main plot shows how $\varepsilon_{\mathrm{I}}$ approaches unity as $G$ increases, following the scaling $1 - \varepsilon_{\mathrm{I}} = O(1/G)$.  
The inset shows the corresponding fluctuation term, which grows $O(G)$.
Therefore, the right-hand side of~\eqref{eq: tradeoff from TUR for feedback cooling} is $O(1)$, which allows the entropy reduction rate to scale as $O(1)$ and remain finite (see also Fig.~\ref{fig:TUR}).

\revision{
\section{Including the effect of harmonic potential}\label{sec: harmonic}

So far, we considered measurement-based feedback applied to a free particle, i.e., without potential.
However, implementations of feedback cooling are realized in systems such as mechanical oscillators or optically levitated nanoparticles.
Analyzing such systems requires incorporating the effect of a harmonic potential into the model.
Moreover, in these systems, feedback control is generally based on the measurement of the position of the system, rather than that of the velocity. We discuss such setting in Sec.~\ref{sec: appendix setup}, and derive a TUR for feedback cooling of a particle trapped in a harmonic potential (see Sec.~\ref{sse: derivation with harmonic} for the derivation):
\begin{align}\label{eq: main trade-off with harmonic potential}
& \left(\frac{\dot{W}_{\mathrm{ext}, u}}{T_u} + \frac{\dot{W}_{\mathrm{ext}, v}}{T_v}\right)\left(1 - \frac{\dot{I}_x}{\dot{I}_{s}}\right)^2 \nonumber \\
&\leq \revision{(\gamma T)^{2}} \|\bm{\nu}_s - \tilde{\bm{\nu}}_s^{\mathrm{irr}}\|_\rho^2 \, \varepsilon_{\mathrm{I}} (1 - \varepsilon_{\mathrm{I}}).
\end{align}
Here, $\dot{I}_{s}$ is the information flow of the system~\eqref{eq: appendix information flow}, $\dot{I}_{x}$ is part of the information flow~\eqref{eq: appendix information flow x} related to the $x$-component of the probability current, and $(T_{u}, \dot{W}_{\mathrm{ext},u})$ and $(T_{v},\dot{W}_{\mathrm{ext},v})$ are the kinetic temperatures and work rates defined along two orthogonal directions~\eqref{eq: appendix Wu}. 

By comparing Eq.~\eqref{eq: main trade-off with harmonic potential} with \eqref{eq: tradeoff from TUR for feedback cooling}, we find that most of the physical intuitions that can be extracted from the TUR holds true even if we add the effect of the harmonic potential. In particular, simaltaneous achievement of the ideal cooling efficiency with finite entropy reduction rate in the asymptotic limit is possible by using the Kalman filter and taking sufficiently large feedback gain (see Sec.~\ref{sse: achievement with harmonic}). 
}

\section{Conclusion}\label{sec: summary}
In this study, we \revision{considered the setup of measurement-based feedback cooling} for underdamped Langevin system and derived the TUR for feedback cooling~\eqref{eq: tradeoff from TUR for feedback cooling}.
To derive the TUR, we introduced a new orthogonal decomposition of the local mean velocity and the partial entropy production rate (Fig.~\ref{fig:orthogonal_decomposition}). 
As a byproduct, we showed that the non-negativity of one of the decomposed entropy productions represents an information thermodynamic cooling limit~\eqref{eq: information flow cooling limit} and recovers the previously known second-law-like inequality~\cite{Horowitz_and_Sandberg_2014} in a special case.
By showing that the information flow can be expressed in the form of the generalized current~\eqref{eq: choice of J}, we obtained a trade-off between the cooling efficiency and entropy reduction rate~\eqref{eq: tradeoff from TUR for feedback cooling}.
\revision{From the obtained trade-off relation, we find that achieving both the ideal cooling efficiency and finite entropy reduction rate is possible by taking the feedback gain to be sufficiently large such that the fluctuation of the reversible mean velocity diverges (see also Fig.~\ref{fig:TUR}).}
While earlier studies provided examples of achieving the ideal efficiency and finite entropy reduction rate~\cite{Horowitz_and_Sandberg_2014, kumasaki2025},
this work proves that such results are consistent with the TUR and clarifies the thermodynamic cost required to achieve such situations. 

\revision{In the case of heat engines, the power-efficiency trade-off relation shows that achieving the Carnot efficiency at finite power requires divergence of an activity-like quantity, which quantifies the effective system-bath coupling strength~\cite{Shiraishi_2016_PhysRevLett.117.190601}. The obtained power-efficiency trade-off relation for feedback cooling~(\ref{eq: tradeoff from TUR for feedback cooling}) has similar mathematical structure, but there are two important differences. First, the definition of the efficiency is different as it includes the kinetic temperature and information flow. Second, the coefficient~(\ref{eq: variation of information flow in tilde}) can be experimentally modified by tuning the feedback gain $G$. For example, the recent nanoparticle experiment using a Kalman filter-based feedback control can tune the feedback gain $G$ to relatively large values to see the asymptotic behavior of the mean occupation number~\cite{Magrini_2021}.}

\revision{We have extended our results by incorporating the effect of harmonic potential~\eqref{eq: main trade-off with harmonic potential}, which is relevant in actual experimental situations, such as levitated nanoparticles in both classical and quantum regime~\cite{Gonzalez_Ballestero_2021}. 
Although our results are obtained by assuming classical systems, an important direction for future work is to extend these results to quantum systems, with the possibility to explore quantum effects such as quantum szueezing~\cite{kamba2025quantum}.}
Developing the TUR for quantum feedback cooling is expected to lead to the clarification of the thermodynamic costs required to achieve the information thermodynamic cooling limit in quantum systems~\cite{kumasaki2025}, and to the thermodynamic characterization of measurement backactions inherent in quantum measurements. 
\revision{Another important direction for future work is to extend the results~\eqref{eq: definition of delta S}-\eqref{eq: information flow cooling limit} and \eqref{eq: main trade-off with harmonic potential} to non-Gaussian distributions, because introducing nonlinear potentials to the system generically makes the probability distribution of the system to be non-Gaussian. }

\begin{acknowledgments}
We are grateful to Kiyotaka Aikawa and Isaac Layton for valuable discussions. 
This work was supported by JST ERATO Grant No. JPMJER2302, Japan.
K.K. and K.T. are supported by World-leading Innovative Graduate Study Program for Materials Research, Information, and Technology (MERIT-WINGS) of the University of Tokyo.
T.S. is supported by JST CREST Grant No. JPMJCR20C1. T.S. is also supported
by Institute of AI and Beyond of the University of Tokyo.
K.F. is supported by JSPS KAKENHI Grant Nos. JP23K13036 and JP24H00831.
\end{acknowledgments}

\appendix 

\section{Setup: including the harmonic potential}\label{sec: appendix setup}



In this appendix, we discuss the setup of feedback cooling of a particle confined in a harmonic potential. 
We consider a single-mode system consisting of the dimensionless position and momentum $(x, p)$.
Let $y$ represent the memory variable corresponding to a noisy measurement of the position of the particle.
We assume that a linear feedback is applied based on an estimate constructed from the measurement outcomes until time $t$.
This situation is modeled by the following set of Langevin equations~\cite{Munakata_2013}:
\begin{align}
\dd x_t &= (\omega\,p_t + a\,u_x(e_t))\,\dd t, \label{eq: ito differential for x with harmonic}\\
\dd p_t &= (-\gamma\,p_t - \omega\,x_t + b\,u_p(e_t))\,\dd t
       + \sqrt{2\,\gamma\,T}\,\dd W_t, \label{eq: ito differential for p with harmonic}\\
\dd y_t &= x_t\, \dd t + \sigma\,\dd W^y_t. \label{eq: ito differential for y with harmonic}
\end{align}
Here, $\omega$ is the frequency of the oscillator, $\gamma$ is the damping coefficient, $T$ is the bath temperature, and $\sigma^2$ is the variance of the measurement noise.
$W_t$ and $W^y_t$ are Wiener processes, which are independent of each other.
The feedback control depends on an estimate $e_t$, which is a vector obtained by compressing the past measurement outcomes $\{\dd y_s\}_{0 \le s \le t}$.
We assume that the feedback forces $u_x(e)$ and $u_p(e)$ in equation~\eqref{eq: ito differential for x with harmonic} depend only on the estimate $e_t$.
While the estimate $e_t$ is generally a non-Markovian process due to its dependence on the past measurement outcomes, we restrict our analysis to the case where the time-evolution equation that describes the updates of $e_t$ is described by a Markov process.
This includes, for example, a direct feedback scheme that uses the current measurement $dy_t$, as well as the Kalman filter.
Specifically, we assume that $e_t$ evolves according to an It\^{o}-type stochastic differential equation:
\begin{align}
    \dd e_t = f(x_t, p_t, e_t)\, \dd t + g(t)\,\dd W_t^e, \label{eq: ito stochastic for estimate}
\end{align}
where the functions $f$ and $g$ are determined by the chosen estimator.

Let $\rho(x,p,e;t)$ denote the probability density function over the extended state space $(x, p, e)$.
Its time evolution is governed by the Fokker-Planck equation:
\begin{equation}
\partial_t \rho(x,p,e;t) = -\bm{\nabla}_{(x,p,e)}\cdot \bm{J}(x,p,e;t),
\end{equation}
where the probability current $\bm{J}_s = (J_x, J_p)^\top$ in the $(x, p)$ subspace is given by
\begin{align}
J_x(x,p,e)&=(\omega p + a u_x(e))\rho,\\
J_p(x,p,e)&=(-\gamma p - \omega x + b u_p(e))\rho - \gamma T \partial_p \rho.
\end{align}
We assume that the initial probability distribution $\rho_{0}(x,p,e)$ is a Gaussian distribution centered at the origin of the phase space (i.e., its first moment vanishes), and the dynamics described above is Gaussian such that it preserves the Gaussianity of the probability distribution during the time-evolution. We further assume that the expectation values of $x$ and $p$ vanish throughout the time-evolution. Note that the example of using the Kalman filter discussed in Sec.~\ref{sse: achievement with harmonic} satisfies these assumptions.

The momentum current $J_p$ can be decomposed into reversible and irreversible components:
\begin{align}
J_p^{\mathrm{irr}}&=-\gamma p\rho - \gamma T \partial_p \rho,\\
J_p^{\mathrm{rev}}&=(-\omega x + b u_p(e))\rho.
\end{align}
In contrast, the position current is entirely reversible: $J_x^{\mathrm{irr}}=0$ and $J_x^{\mathrm{rev}}=J_x$. 
This decomposition reflects the assumption that the feedback forces are derived from a linear Hamiltonian in $(x,p)$ and thus contribute only to reversible dynamics.
We write the reversible and irreversible current vectors in the subspace $(x, p)$ as $\bm{J}^{\mathrm{irr}}_s = (0, J_p^{\mathrm{irr}})^\top$ and $\bm{J}^{\mathrm{rev}}_s = (J_x^{\mathrm{rev}}, J_p^{\mathrm{rev}})^\top$, respectively.
The average energy of the system is defined as
\begin{equation}
    E \coloneqq \int \left(\frac{1}{2}\omega p^2 + \frac{1}{2}\omega x^2\right) \rho(x, p, e)\, \dd x \dd p \dd e.
\end{equation}
From the time derivative of this energy, we define the heat current as
\begin{equation}
    \dot{Q} \coloneqq \int \bm{s}^\top \bm{J}_s^{\mathrm{irr}}\, \dd x \dd p \dd e,
\end{equation}
and the work rate due to feedback as
\begin{equation}
    \dot{W}_{\mathrm{ext}} \coloneqq \int \bm{s}^\top \bm{J}_s^{\mathrm{rev}}\, \dd x \dd p \dd e,
\end{equation}
where $\bm{s} = [x, p]^\top$.
These definitions satisfy the first law of thermodynamics in the form
$
\dot{E} = -\dot{W}_{\mathrm{ext}} + \dot{Q}.
$

\section{Orthogonal decomposition of the partial entropy production}\label{sse: orthogonal decomp with harmonic}
As in the case without a harmonic potential, we introduce an orthogonal decomposition in this setting.
In this section, we derive an information thermodynamic cooling limit based on this decomposition.
The partial entropy production rate is defined in analogy with the one-dimensional case as
\begin{align}
\dot{\Sigma}_s \coloneqq \frac{1}{\gamma T} \int \|\bm{\nu}_s^{\mathrm{irr}}(x,p,e)\|^2 \rho(x,p,e)\, \dd x \dd p \dd e,
\end{align}
where the irreversible local mean velocity is given by $\bm{\nu}_s^{\mathrm{irr}}(x,p,e) \coloneqq \bm{J}_s^{\mathrm{irr}}(x,p,e) / \rho(x,p,e)$.
We define an inner product for vector functions $\bm{f}(x,p,e)$ and $\bm{g}(x,p,e)$ as
\begin{align}
\langle \bm{f}, \bm{g} \rangle_\rho \coloneqq (1/\gamma T) \int \bm{f}^\top(x,p,e) \bm{g}(x,p,e) \rho(x,p,e)\, \dd x \dd p \dd e .
\end{align}
Using the associated norm, the partial entropy production rate can be written compactly as $\dot{\Sigma}_s = \|\bm{\nu}_s^{\mathrm{irr}}\|_\rho^2$.

Next, we introduce an orthogonal decomposition of $\bm{\nu}_s^{\mathrm{irr}}$.
Let $\tilde\rho(x,p) \coloneqq \int \rho(x,p,e)\, \dd e$ be the marginal distribution of the system, and define the marginal irreversible probability current and corresponding local mean velocity as
\begin{align}
\tilde{\bm{J}}_s^{\mathrm{irr}}(x,p) &\coloneqq \int \bm{J}_s^{\mathrm{irr}}(x,p,e)\, \dd e, \\
\tilde{\bm{\nu}}_s^{\mathrm{irr}}(x,p) &\coloneqq \tilde{\bm{J}}_s^{\mathrm{irr}}(x,p) / \tilde\rho(x,p).
\end{align}
The residual local mean velocity is then defined as
 $
\delta \bm{\nu}_s^{\mathrm{irr}}(x,p,e) \coloneqq \bm{\nu}_s^{\mathrm{irr}}(x,p,e) - \tilde{\bm{\nu}}_s^{\mathrm{irr}}(x,p).
$
By construction, this decomposition satisfies the orthogonality condition:
$
\langle \tilde{\bm{\nu}}_s^{\mathrm{irr}}, \delta \bm{\nu}_s^{\mathrm{irr}} \rangle_\rho = 0.
$
As a result, the partial entropy production rate can be decomposed as follows
\begin{equation}
\dot{\Sigma}_s
= \|\bm{\nu}_s^{\mathrm{irr}}\|^2_\rho
= \|\tilde{\bm{\nu}}_s^{\mathrm{irr}}\|^2_\rho
+ \|\delta \bm{\nu}_s^{\mathrm{irr}}\|^2_\rho
\eqcolon\dot{\Sigma}_{\tilde{\nu}^{\mathrm{irr}}} + \dot{\Sigma}_{\delta \nu^{\mathrm{irr}}}.
\end{equation}

In what follows, we show that the non-negativity of $\dot{\Sigma}_{\delta \nu^{\mathrm{irr}}}$ leads to an information thermodynamic cooling limit.
We define the information flow and its reversible and irreversible components as
\begin{align} \label{eq: appendix information flow}
\dot{I}_{s} &\coloneqq \int \bm{J}_{\mathrm{s}}^\top \nabla_{\mathrm{s}} \ln \frac{\rho(x,p,e)}{\tilde \rho(x,p)}\, \dd x \dd p \dd e =\dot{I}_{s}^{\mathrm{irr}}+\dot{I}_{s}^{\mathrm{rev}} ,\\
\dot{I}_{s}^{\mathrm{irr}} &\coloneqq \int \bm{J}_{\mathrm{s}}^{\mathrm{irr}\top} \nabla_{\mathrm{s}} \ln \frac{\rho(x,p,e)}{\tilde \rho(x,p)}\, \dd x \dd p \dd e,\\
\dot{I}_{s}^{\mathrm{rev}} &\coloneqq \int \bm{J}_{\mathrm{s}}^{\mathrm{rev}\top} \nabla_{\mathrm{s}} \ln \frac{\rho(x,p,e)}{\tilde \rho(x,p)}\, \dd x \dd p \dd e.
\end{align}
A direct computation yields the following relation:
\begin{align}
     -\dot{I}_{s}^{\mathrm{irr}}=\dot{\Sigma}_{\delta \nu^{\mathrm{irr}}} \geq 0.
\end{align}
Therefore, the non-negativity of $\dot{\Sigma}_{\delta \nu^{\mathrm{irr}}}$ leads to the relation
\begin{align}
-\dot{I}_{s}^{\mathrm{rev}} \le -\dot{I}_{s}. \label{eq: inequality between I_s^rev and I_s}
\end{align}
We proceed to rewrite the reversible information flow.
Let $\Xi \coloneqq \langle (x, p)^\top (x, p) \rangle$ be the covariance matrix of $(x,p)$, and let $R$ be the orthogonal matrix that diagonalizes $\Xi$.
Define the new coordinates $(u, v)^\top \coloneqq R^\top (x, p)^\top$, and define a reversible probability flow in the new coordinates as
\begin{equation}
\bm{J}_s^{\mathrm{rev}'}(u,v,e) \coloneqq R^\top \bm{J}_s^{\mathrm{rev}}(x,p,e).
\end{equation}
Then, the work rates in the eigen directions are defined as
\begin{align}\label{eq: appendix Wu}
\dot{W}_{\mathrm{ext}, u} &\coloneqq \int u J_u^{\mathrm{rev}'}(u,v,e)\, \dd u \dd v \dd e, \\
\dot{W}_{\mathrm{ext}, v} &\coloneqq \int v J_v^{\mathrm{rev}'}(u,v,e)\, \dd u \dd v \dd e.
\end{align}
These satisfy $\dot{W}_{\mathrm{ext}} = \dot{W}_{\mathrm{ext}, u} + \dot{W}_{\mathrm{ext}, v}$, giving a decomposition of the total work rate.
We define the kinetic temperatures in each direction as
$T_u \coloneqq \omega \sigma_{uu}$ and $T_v \coloneqq \omega \sigma_{vv}$, 
where $\sigma_{uu}$ and $\sigma_{vv}$ are the corresponding variances.
Using these definitions, the reversible part of the information flow becomes
\begin{align}
-\dot{I}_{s}^{\mathrm{rev}}
&= -\int \bm{J}_{\mathrm{s}}^{\mathrm{rev}' \top} \nabla_{\mathrm{s}}' \ln \rho(u,v)\, \dd u \dd v \dd e \notag \\
&=  \frac{\dot{W}_{\mathrm{ext}, u}}{T_u} + \frac{\dot{W}_{\mathrm{ext}, v}}{T_v}, \label{eq: reversible information flow}
\end{align}
where $\nabla_{\mathrm{s}}' \coloneqq R^\top \nabla_{\mathrm{s}}$.

Combining the results~\eqref{eq: inequality between I_s^rev and I_s} and~\eqref{eq: reversible information flow}, we obtain the information thermodynamic cooling limit for an underdamped system with harmonic potential:
\begin{equation}
\frac{\dot{W}_{\mathrm{ext}, u}}{T_u} + \frac{\dot{W}_{\mathrm{ext}, v}}{T_v}\;\le\;- \dot{I}_{s}. \label{eq: information flow cooling limit with harmonic potential}
\end{equation}
The left-hand side \revision{is a generalization of the entropy reduction rate $\dot{W}_{\rm ext}/T_{\rm kin}$ to the case where the particle is trapped in a harmonic potential.}
We define the cooling efficiency with respect to information flow as
\begin{align}
\varepsilon_{\mathrm{I}} \coloneqq \left|\frac{\dot{W}_{\mathrm{ext}, u}/T_u + \dot{W}_{\mathrm{ext}, v}/T_v}{\dot{I}_{s}}\right|,
\end{align}
which satisfies $\varepsilon_{\mathrm{I}} \le 1$ when $\dot{W}_{\mathrm{ext}, u}/T_u + \dot{W}_{\mathrm{ext}, v}/T_v\geq 0$, i.e., when the energy of the system is reduced by feedback.
Finally, using the bound $-\dot{I}_{s} \le \dot{I}_{\mathrm{TE}}$ between the information flow and transfer entropy \revision{in the steady-state}, we obtain
\begin{align}
\frac{\dot{W}_{\mathrm{ext}, u}}{T_u}
+ \frac{\dot{W}_{\mathrm{ext}, v}}{T_v}
\;\le\;
 \dot{I}_{\mathrm{TE}}.
\end{align}
This result is the classical counterpart of the information thermodynamic cooling limit  for quantum systems under harmonic potential~\cite{kumasaki2025}.

\section{Derivation of the TUR}\label{sse: derivation with harmonic}
In this section, we derive the short-time TUR that characterizes the fundamental limit of feedback cooling, and represents a trade-off relation between the cooling efficiency and entropy reduction rate.
As in the case without a potential, we define the generalized current associated with the irreversible flow as
$
\mathcal{J}_{\nu^{\rm irr}}^{\bm{w}} \coloneqq \gamma T \ev{\bm{w}, \bm{\nu}_s^{\mathrm{irr}}}_\rho,
$
where $\bm{w}(x,p,e)$ is an arbitrary weight function vector.
Using the orthogonal decomposition, this generalized irreversible current can be decomposed as
\begin{align}
\mathcal{J}_{\nu^{\rm irr}}^{\bm{w}}
&=  \mathcal{J}_{\delta\nu^{\rm irr}}^{\bm{w}} + \mathcal{J}_{\tilde{\nu}^{\rm irr}}^{\bm{w}},\\
\mathcal{J}_{\delta\nu^{\rm irr}}^{\bm{w}}
&\coloneqq \gamma T\ev{\bm{w}, \delta\bm{\nu}_s^{\mathrm{irr}}}_\rho, \label{eq:current_delta_nu}\\
\mathcal{J}_{\tilde{\nu}^{\rm irr}}^{\bm{w}}
&\coloneqq \gamma T\ev{\bm{w}, \widetilde{\bm{\nu}}_s^{\mathrm{irr}}}_\rho.
\end{align}
From the Cauchy-Schwarz inequality, these components satisfy the following short-time TURs:
\begin{align}
\left(\mathcal{J}_{\nu^{\rm irr}}^{\bm{w}}\right)^2 &\leq \revision{(\gamma T)^{2}}\left\|\bm{w}\right\|_\rho^2\dot{\Sigma}_s, \label{eq: short time TUR for partial entropy production with harmonic potential} \\
\left(\mathcal{J}_{\delta\nu^{\rm irr}}^{\bm{w}}\right)^2 &\leq \revision{(\gamma T)^{2}}\left\|\bm{w}\right\|_\rho^2\dot{\Sigma}_{\delta \nu^{\mathrm{irr}}}, \label{eq: short time TUR for feedback cooling with harmonic potential} \\
\left(\mathcal{J}_{\tilde{\nu}^{\rm irr}}^{\bm{w}}\right)^2 &\leq \revision{(\gamma T)^{2}}\left\|\bm{w}\right\|_\rho^2\dot{\Sigma}_{\tilde{\nu}^{\mathrm{irr}}}.
\label{eq: short time TUR for marginal entropy production with harmonic potential}
\end{align}
As in the free particle case, the non-negativity of $\dot{\Sigma}_{\delta \nu}$ implies the inequality~\eqref{eq: information flow cooling limit with harmonic potential}.
Therefore, Eq.~\eqref{eq: short time TUR for feedback cooling with harmonic potential} can be regarded as a short-time TUR corresponding to the cooling limit.

We now derive a trade-off between the cooling efficiency and entropy reduction rate from equation~\eqref{eq: short time TUR for feedback cooling with harmonic potential}.
We define information flow along the direction of $x$ and $p$ as
\begin{align} \label{eq: appendix information flow x}
\dot{I}_{x} &\coloneqq \int J_{x} \partial_x \ln \frac{\rho(x,p,e)}{\tilde \rho(x,p)}\, \dd x \dd p \dd e \\
\dot{I}_{p} &\coloneqq \int J_{p} \partial_p \ln \frac{\rho(x,p,e)}{\tilde \rho(x,p)}\, \dd x \dd p \dd e,
\end{align}
with
\begin{align}
    \dot{I}_{s}=\dot{I}_{x} + \dot{I}_{p}. 
\end{align}
By choosing $\bm{w} = \revision{\bm{\nu}_s} - \tilde{\bm{\nu}}_s^{\mathrm{irr}}$, we find\revision{
$
\mathcal{J}_{\delta\nu^{\rm irr}}^{\bm{w}} = -\dot{I}_p.
$}
Substituting this into equation~\eqref{eq: short time TUR for feedback cooling with harmonic potential} and assuming the cooling condition $\dot{W}_{\mathrm{ext}, u}/T_u + \dot{W}_{\mathrm{ext}, v}/T_v\geq 0$ yields
\begin{align}\label{eq: trade-off with harmonic potential}
& \left(\frac{\dot{W}_{\mathrm{ext}, u}}{T_u} + \frac{\dot{W}_{\mathrm{ext}, v}}{T_v}\right)\left(1 - \frac{\dot{I}_x}{\dot{I}_{s}}\right)^2 \nonumber \\
&\leq \revision{(\gamma T)^{2}} \|\bm{\nu}_s - \tilde{\bm{\nu}}_s^{\mathrm{irr}}\|_\rho^2 \, \varepsilon_{\mathrm{I}} (1 - \varepsilon_{\mathrm{I}}),
\end{align}
which shows a refined trade-off compared to~(15) in the main text obtained for the free particle case
due to the additional factor $\left(1 - \dot{I}_x / \dot{I}_{s}\right)^2$ on the left-hand side.
To achieve both finite entropy reduction rate and maximal efficiency $\varepsilon_{\mathrm{I}} = 1$ simultaneously,
the inequality implies that either $\|\bm{\nu}_s - \tilde{\bm{\nu}}_s^{\mathrm{irr}}\|_\rho^2 \rightarrow \infty$ or $\left(1 - \dot{I}_x / \dot{I}_{s}\right)^2 \rightarrow 0$ is needed.
In feedback cooling scenarios, $\dot{I}_p$ typically remains finite, hence $\left(1 - \dot{I}_x / \dot{I}_{s}\right)^2$ does not vanish.
Therefore, the divergence of the fluctuation term $\|\bm{\nu}_s - \tilde{\bm{\nu}}_s^{\mathrm{irr}}\|_\rho^2$ is necessary.

\revision{
We breifely mention how the result~(\ref{eq: trade-off with harmonic potential}) is related to Eq.~\eqref{eq: tradeoff from TUR for feedback cooling} valid for free particle case. When the harmonic potential is absent and when we only apply the feedback force $b u_{p}(e_{t})$ (that is, $au_{x}(e)=0$), the Langevin equation~\eqref{eq: ito differential for x with harmonic} and \eqref{eq: ito differential for p with harmonic} is consistent with Eq.~\eqref{eq: LE sys}, although the way of modeling the measurement is different in these two cases. Nevertheless, one can show that  
\begin{align}
    \dot{I}_{x} = \int \omega p \rho(x,p,e) \partial_{x} \ln \frac{ \rho(x,p,e)}{\tilde{\rho}(x,p)} \, \dd x \dd p \dd e = 0,
\end{align}
by assuming $au_{x}(e)=0$. Therefore, Eq.~\eqref{eq: trade-off with harmonic potential} reads
\begin{align} \label{eq: trade-off xzero}
\left(\frac{\dot{W}_{\mathrm{ext}, u}}{T_u} + \frac{\dot{W}_{\mathrm{ext}, v}}{T_v}\right) \leq (\gamma T)^{2} \|\bm{\nu}_s - \tilde{\bm{\nu}}_s^{\mathrm{irr}}\|_\rho^2 \, \varepsilon_{\mathrm{I}} (1 - \varepsilon_{\mathrm{I}}),
\end{align}
which is very similar to Eq.~\eqref{eq: tradeoff from TUR for feedback cooling}, but there exists a slight difference because we only consider the velocity degrees of freedom of the system in Eq.~\eqref{eq: tradeoff from TUR for feedback cooling}, whereas we consider both the position and momentum degrees of freedom in Eq.~\ref{eq: trade-off xzero}. Therefore, $\dot{I}_{v}\neq \dot{I}_{s}$ in general, since $\dot{I}_{v}$ quantifies the change in the correlation between $v$ and $y$ arising from the system dynamics, whereas $\dot{I}_{s}$ quantifies that between $(x,p)$ and $e$. We further note that instead of just one kinetic temperature $T_{\rm kin}$ in Eq.~\eqref{eq: tradeoff from TUR for feedback cooling}, we have two kinetic temperatures $T_{u}$ and $T_{v}$ in Eq.~\eqref{eq: trade-off with harmonic potential}.   
}

\begin{figure}[tbp]
\centering
\includegraphics[width=0.95\linewidth]{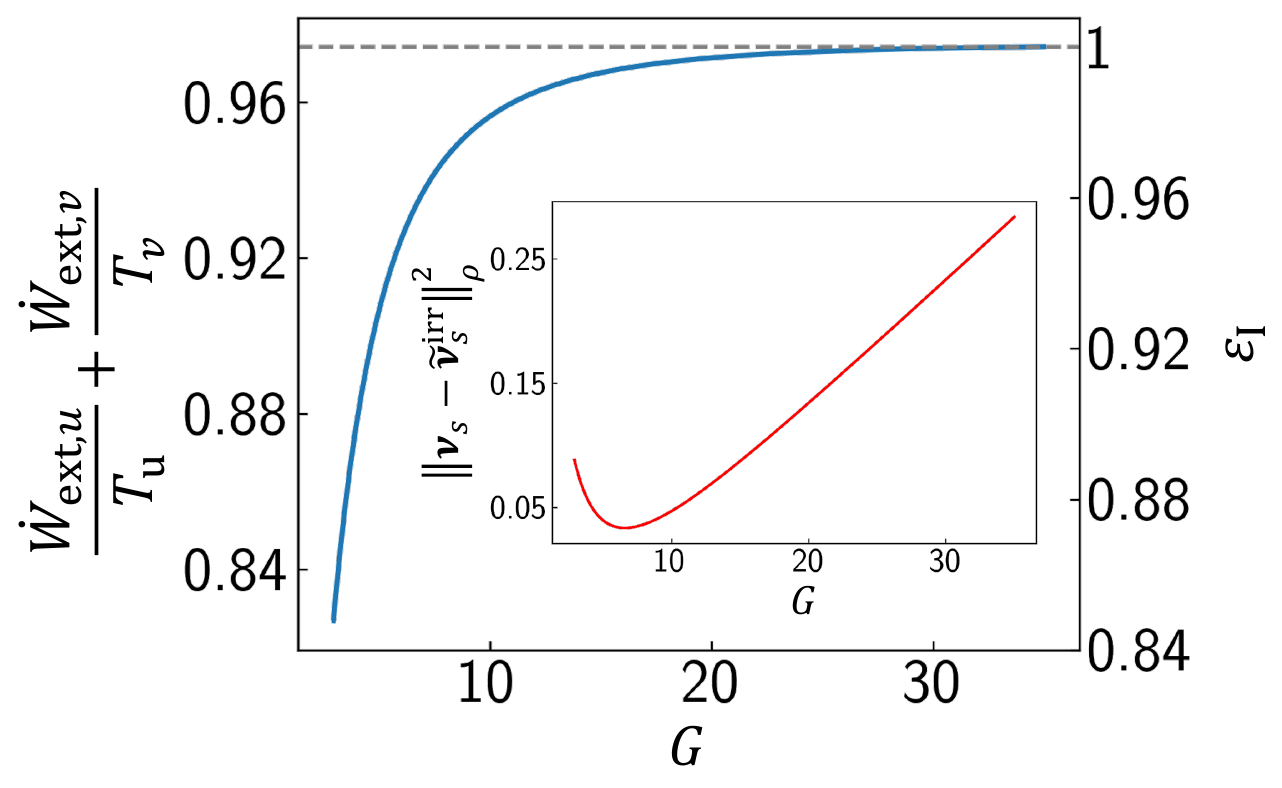}
\caption{Behavior of the cooling efficiency and entropy reduction rate as functions of the feedback gain $G$.
The main plot shows the entropy reduction rate $\dot{W}_{\mathrm{ext}, u}/T_u + \dot{W}_{\mathrm{ext}, v}/T_v$ (left axis) and the cooling efficiency $\varepsilon_{\mathrm{I}}$ (right axis)\revision{, where the two curves overlap with each other because the information flow $\dot{I}_{s}$ does not depend on $G$ in this setup.}
As $G$ increases, $\varepsilon_{\mathrm{I}}$ asymptotically approaches unity, while the fluctuation associated with the entropy reduction rate (shown in the inset) diverges linearly.
This divergence clarifies how asymptotically achieving finite entropy reduction rate at maximum efficiency is possible from the viewpoint of the trade-off relation Eq.~\eqref{eq: trade-off with harmonic potential}. The parameters are: $\omega=1$, $\gamma=10$, $T=1$, and $\sigma=0.2$.}
\label{fig: achieving max efficiency harmonic}
\end{figure}

\section{Simultaneous achievement of maximum efficiency and finite entropy reduction rate}\label{sse: achievement with harmonic}
In this section, we numerically demonstrate that when the Kalman filter is used as the estimator,
the system asymptotically achieves maximum efficiency and finite entropy reduction rate in the limit of infinite feedback gain.
We further show that $\|\bm{\nu}_s - \tilde{\bm{\nu}}_s^{\mathrm{irr}}\|_\rho^2$ diverges in accordance with the trade-off relation~\eqref{eq: trade-off with harmonic potential}.
We begin by describing the Kalman filter dynamics corresponding to equations~\eqref{eq: ito differential for x with harmonic}-\eqref{eq: ito differential for y with harmonic}.
The estimated state $e_t = [\hat{x}, \hat{p}]^\top$ evolves according to~\cite{simon2006optimal}:
\begin{align}
\dd \hat{x} &= (\omega \hat{p} + a u_x)\,\dd t + \frac{\hat{V}_{xx}}{\sigma^2}(\dd y - \hat{x}\,\dd t), \label{eq: kalman x}\\
\dd \hat{p} &= (-\gamma \hat{p} - \omega \hat{x} + b u_p)\,\dd t + \frac{\hat{V}_{xp}}{\sigma^2}(\dd y - \hat{x}\,\dd t), \label{eq: kalman p}
\end{align}
where $\hat{\bm{V}}$ is the estimation error covariance matrix, determined by the Riccati equation:
\begin{align}
\frac{\dd}{\dd t} \hat{V}_{xx} &= 2\omega \hat{V}_{xp} - \frac{\hat{V}_{xx}^2}{\sigma^2}, \\
\frac{\dd}{\dd t} \hat{V}_{pp} &= -2\omega \hat{V}_{xp} - 2\gamma \hat{V}_{pp} + 2\gamma T - \frac{\hat{V}_{xp}^2}{\sigma^2}, \\
\frac{\dd}{\dd t} \hat{V}_{xp} &= -\omega \hat{V}_{xx} + \omega \hat{V}_{pp} - \gamma \hat{V}_{xp} - \frac{\hat{V}_{xx} \hat{V}_{xp}}{\sigma^2}.
\end{align}
Since this is a closed set of equations, the time evolution of $\hat{\bm{V}}$ can be obtained in advance.
Therefore, the estimation dynamics~\eqref{eq: kalman x}-\eqref{eq: kalman p} take the form of~\eqref{eq: ito stochastic for estimate},
and the full theory developed in the previous sections applies to the extended state space $[x, p, \hat{x}, \hat{p}]^\top$.

We focus on the case where the feedback is set as $u_x = -\hat{x}$ and $u_p = -\hat{p}$,
and take the feedback gains to be equal: $a = b = G$.
Under this setup, we numerically calculate the behavior of $\dot{W}_{\mathrm{ext}, u}/T_u + \dot{W}_{\mathrm{ext}, v}/T_v$, $\varepsilon_{\mathrm{I}}$, and $\|\bm{\nu}_s - \tilde{\bm{\nu}}_s^{\mathrm{irr}}\|_\rho^2$ as functions of $G$.
These are shown in Figure~\ref{fig: achieving max efficiency harmonic}.
The main plot shows that $\varepsilon_{\mathrm{I}} \to 1$ as $G$ increases,
with $1 - \varepsilon_{\mathrm{I}} = O(1/G)$. The inset confirms that $\|\bm{\nu}_s - \tilde{\bm{\nu}}_s^{\mathrm{irr}}\|_\rho^2$ diverges linearly in the limit of large $G$.
Hence, the right-hand side of the trade-off inequality~\eqref{eq: trade-off with harmonic potential} remains finite,
and the system achieves feedback cooling with finite entropy reduction rate at maximum efficiency.


\begin{thebibliography}{51}%
\makeatletter
\providecommand \@ifxundefined [1]{%
 \@ifx{#1\undefined}
}%
\providecommand \@ifnum [1]{%
 \ifnum #1\expandafter \@firstoftwo
 \else \expandafter \@secondoftwo
 \fi
}%
\providecommand \@ifx [1]{%
 \ifx #1\expandafter \@firstoftwo
 \else \expandafter \@secondoftwo
 \fi
}%
\providecommand \natexlab [1]{#1}%
\providecommand \enquote  [1]{``#1''}%
\providecommand \bibnamefont  [1]{#1}%
\providecommand \bibfnamefont [1]{#1}%
\providecommand \citenamefont [1]{#1}%
\providecommand \href@noop [0]{\@secondoftwo}%
\providecommand \href [0]{\begingroup \@sanitize@url \@href}%
\providecommand \@href[1]{\@@startlink{#1}\@@href}%
\providecommand \@@href[1]{\endgroup#1\@@endlink}%
\providecommand \@sanitize@url [0]{\catcode `\\12\catcode `\$12\catcode `\&12\catcode `\#12\catcode `\^12\catcode `\_12\catcode `\%12\relax}%
\providecommand \@@startlink[1]{}%
\providecommand \@@endlink[0]{}%
\providecommand \url  [0]{\begingroup\@sanitize@url \@url }%
\providecommand \@url [1]{\endgroup\@href {#1}{\urlprefix }}%
\providecommand \urlprefix  [0]{URL }%
\providecommand \Eprint [0]{\href }%
\providecommand \doibase [0]{https://doi.org/}%
\providecommand \selectlanguage [0]{\@gobble}%
\providecommand \bibinfo  [0]{\@secondoftwo}%
\providecommand \bibfield  [0]{\@secondoftwo}%
\providecommand \translation [1]{[#1]}%
\providecommand \BibitemOpen [0]{}%
\providecommand \bibitemStop [0]{}%
\providecommand \bibitemNoStop [0]{.\EOS\space}%
\providecommand \EOS [0]{\spacefactor3000\relax}%
\providecommand \BibitemShut  [1]{\csname bibitem#1\endcsname}%
\let\auto@bib@innerbib\@empty
\bibitem [{\citenamefont {Sekimoto}(2010)}]{sekimoto2010stochastic}%
  \BibitemOpen
  \bibfield  {author} {\bibinfo {author} {\bibfnamefont {K.}~\bibnamefont {Sekimoto}},\ }\href@noop {} {\emph {\bibinfo {title} {Stochastic energetics}}}\ (\bibinfo  {publisher} {Springer},\ \bibinfo {year} {2010})\BibitemShut {NoStop}%
\bibitem [{\citenamefont {Esposito}\ \emph {et~al.}(2009)\citenamefont {Esposito}, \citenamefont {Harbola},\ and\ \citenamefont {Mukamel}}]{RevModPhys.81.1665}%
  \BibitemOpen
  \bibfield  {author} {\bibinfo {author} {\bibfnamefont {M.}~\bibnamefont {Esposito}}, \bibinfo {author} {\bibfnamefont {U.}~\bibnamefont {Harbola}},\ and\ \bibinfo {author} {\bibfnamefont {S.}~\bibnamefont {Mukamel}},\ }\bibfield  {title} {\bibinfo {title} {Nonequilibrium fluctuations, fluctuation theorems, and counting statistics in quantum systems},\ }\href {https://doi.org/10.1103/RevModPhys.81.1665} {\bibfield  {journal} {\bibinfo  {journal} {Rev. Mod. Phys.}\ }\textbf {\bibinfo {volume} {81}},\ \bibinfo {pages} {1665} (\bibinfo {year} {2009})}\BibitemShut {NoStop}%
\bibitem [{\citenamefont {Campisi}\ \emph {et~al.}(2011)\citenamefont {Campisi}, \citenamefont {H\"anggi},\ and\ \citenamefont {Talkner}}]{RevModPhys.83.771}%
  \BibitemOpen
  \bibfield  {author} {\bibinfo {author} {\bibfnamefont {M.}~\bibnamefont {Campisi}}, \bibinfo {author} {\bibfnamefont {P.}~\bibnamefont {H\"anggi}},\ and\ \bibinfo {author} {\bibfnamefont {P.}~\bibnamefont {Talkner}},\ }\bibfield  {title} {\bibinfo {title} {Colloquium: Quantum fluctuation relations: Foundations and applications},\ }\href {https://doi.org/10.1103/RevModPhys.83.771} {\bibfield  {journal} {\bibinfo  {journal} {Rev. Mod. Phys.}\ }\textbf {\bibinfo {volume} {83}},\ \bibinfo {pages} {771} (\bibinfo {year} {2011})}\BibitemShut {NoStop}%
\bibitem [{\citenamefont {Seifert}(2012)}]{Seifert_2012}%
  \BibitemOpen
  \bibfield  {author} {\bibinfo {author} {\bibfnamefont {U.}~\bibnamefont {Seifert}},\ }\bibfield  {title} {\bibinfo {title} {Stochastic thermodynamics, fluctuation theorems and molecular machines},\ }\href {https://doi.org/10.1088/0034-4885/75/12/126001} {\bibfield  {journal} {\bibinfo  {journal} {Reports on Progress in Physics}\ }\textbf {\bibinfo {volume} {75}},\ \bibinfo {pages} {126001} (\bibinfo {year} {2012})}\BibitemShut {NoStop}%
\bibitem [{\citenamefont {Ciliberto}(2017)}]{PhysRevX.7.021051}%
  \BibitemOpen
  \bibfield  {author} {\bibinfo {author} {\bibfnamefont {S.}~\bibnamefont {Ciliberto}},\ }\bibfield  {title} {\bibinfo {title} {Experiments in stochastic thermodynamics: Short history and perspectives},\ }\href {https://doi.org/10.1103/PhysRevX.7.021051} {\bibfield  {journal} {\bibinfo  {journal} {Phys. Rev. X}\ }\textbf {\bibinfo {volume} {7}},\ \bibinfo {pages} {021051} (\bibinfo {year} {2017})}\BibitemShut {NoStop}%
\bibitem [{\citenamefont {Funo}\ \emph {et~al.}(2018)\citenamefont {Funo}, \citenamefont {Ueda},\ and\ \citenamefont {Sagawa}}]{Funo2018}%
  \BibitemOpen
  \bibfield  {author} {\bibinfo {author} {\bibfnamefont {K.}~\bibnamefont {Funo}}, \bibinfo {author} {\bibfnamefont {M.}~\bibnamefont {Ueda}},\ and\ \bibinfo {author} {\bibfnamefont {T.}~\bibnamefont {Sagawa}},\ }\bibinfo {title} {Quantum fluctuation theorems},\ in\ \href {https://doi.org/10.1007/978-3-319-99046-0_10} {\emph {\bibinfo {booktitle} {Thermodynamics in the Quantum Regime: Fundamental Aspects and New Directions}}},\ \bibinfo {editor} {edited by\ \bibinfo {editor} {\bibfnamefont {F.}~\bibnamefont {Binder}}, \bibinfo {editor} {\bibfnamefont {L.~A.}\ \bibnamefont {Correa}}, \bibinfo {editor} {\bibfnamefont {C.}~\bibnamefont {Gogolin}}, \bibinfo {editor} {\bibfnamefont {J.}~\bibnamefont {Anders}},\ and\ \bibinfo {editor} {\bibfnamefont {G.}~\bibnamefont {Adesso}}}\ (\bibinfo  {publisher} {Springer International Publishing},\ \bibinfo {address} {Cham},\ \bibinfo {year} {2018})\ pp.\ \bibinfo {pages} {249--273}\BibitemShut {NoStop}%
\bibitem [{\citenamefont {Landi}\ and\ \citenamefont {Paternostro}(2021)}]{RevModPhys.93.035008}%
  \BibitemOpen
  \bibfield  {author} {\bibinfo {author} {\bibfnamefont {G.~T.}\ \bibnamefont {Landi}}\ and\ \bibinfo {author} {\bibfnamefont {M.}~\bibnamefont {Paternostro}},\ }\bibfield  {title} {\bibinfo {title} {Irreversible entropy production: From classical to quantum},\ }\href {https://doi.org/10.1103/RevModPhys.93.035008} {\bibfield  {journal} {\bibinfo  {journal} {Rev. Mod. Phys.}\ }\textbf {\bibinfo {volume} {93}},\ \bibinfo {pages} {035008} (\bibinfo {year} {2021})}\BibitemShut {NoStop}%
\bibitem [{\citenamefont {Parrondo}\ \emph {et~al.}(2015)\citenamefont {Parrondo}, \citenamefont {Horowitz},\ and\ \citenamefont {Sagawa}}]{Parrondo_Horowitz_Sagawa_2015}%
  \BibitemOpen
  \bibfield  {author} {\bibinfo {author} {\bibfnamefont {J.~M.}\ \bibnamefont {Parrondo}}, \bibinfo {author} {\bibfnamefont {J.}~\bibnamefont {Horowitz}},\ and\ \bibinfo {author} {\bibfnamefont {T.}~\bibnamefont {Sagawa}},\ }\bibfield  {title} {\bibinfo {title} {Thermodynamics of information},\ }\href {https://doi.org/10.1038/nphys3230} {\bibfield  {journal} {\bibinfo  {journal} {Nature Physics}\ }\textbf {\bibinfo {volume} {11}},\ \bibinfo {pages} {131} (\bibinfo {year} {2015})}\BibitemShut {NoStop}%
\bibitem [{\citenamefont {Sagawa}\ and\ \citenamefont {Ueda}(2008)}]{PhysRevLett.100.080403}%
  \BibitemOpen
  \bibfield  {author} {\bibinfo {author} {\bibfnamefont {T.}~\bibnamefont {Sagawa}}\ and\ \bibinfo {author} {\bibfnamefont {M.}~\bibnamefont {Ueda}},\ }\bibfield  {title} {\bibinfo {title} {Second law of thermodynamics with discrete quantum feedback control},\ }\href {https://doi.org/10.1103/PhysRevLett.100.080403} {\bibfield  {journal} {\bibinfo  {journal} {Phys. Rev. Lett.}\ }\textbf {\bibinfo {volume} {100}},\ \bibinfo {pages} {080403} (\bibinfo {year} {2008})}\BibitemShut {NoStop}%
\bibitem [{\citenamefont {Horowitz}\ and\ \citenamefont {Esposito}(2014)}]{PhysRevX.4.031015}%
  \BibitemOpen
  \bibfield  {author} {\bibinfo {author} {\bibfnamefont {J.~M.}\ \bibnamefont {Horowitz}}\ and\ \bibinfo {author} {\bibfnamefont {M.}~\bibnamefont {Esposito}},\ }\bibfield  {title} {\bibinfo {title} {Thermodynamics with continuous information flow},\ }\href {https://doi.org/10.1103/PhysRevX.4.031015} {\bibfield  {journal} {\bibinfo  {journal} {Phys. Rev. X}\ }\textbf {\bibinfo {volume} {4}},\ \bibinfo {pages} {031015} (\bibinfo {year} {2014})}\BibitemShut {NoStop}%
\bibitem [{\citenamefont {Yada}\ \emph {et~al.}(2022)\citenamefont {Yada}, \citenamefont {Yoshioka},\ and\ \citenamefont {Sagawa}}]{PhysRevLett.128.170601}%
  \BibitemOpen
  \bibfield  {author} {\bibinfo {author} {\bibfnamefont {T.}~\bibnamefont {Yada}}, \bibinfo {author} {\bibfnamefont {N.}~\bibnamefont {Yoshioka}},\ and\ \bibinfo {author} {\bibfnamefont {T.}~\bibnamefont {Sagawa}},\ }\bibfield  {title} {\bibinfo {title} {Quantum fluctuation theorem under quantum jumps with continuous measurement and feedback},\ }\href {https://doi.org/10.1103/PhysRevLett.128.170601} {\bibfield  {journal} {\bibinfo  {journal} {Phys. Rev. Lett.}\ }\textbf {\bibinfo {volume} {128}},\ \bibinfo {pages} {170601} (\bibinfo {year} {2022})}\BibitemShut {NoStop}%
\bibitem [{\citenamefont {Ptaszy\ifmmode~\acute{n}\else \'{n}\fi{}ski}\ and\ \citenamefont {Esposito}(2019)}]{PhysRevLett.122.150603}%
  \BibitemOpen
  \bibfield  {author} {\bibinfo {author} {\bibfnamefont {K.}~\bibnamefont {Ptaszy\ifmmode~\acute{n}\else \'{n}\fi{}ski}}\ and\ \bibinfo {author} {\bibfnamefont {M.}~\bibnamefont {Esposito}},\ }\bibfield  {title} {\bibinfo {title} {Thermodynamics of quantum information flows},\ }\href {https://doi.org/10.1103/PhysRevLett.122.150603} {\bibfield  {journal} {\bibinfo  {journal} {Phys. Rev. Lett.}\ }\textbf {\bibinfo {volume} {122}},\ \bibinfo {pages} {150603} (\bibinfo {year} {2019})}\BibitemShut {NoStop}%
\bibitem [{\citenamefont {Yada}\ \emph {et~al.}(2025)\citenamefont {Yada}, \citenamefont {Stas}, \citenamefont {Suleymanzade}, \citenamefont {Knall}, \citenamefont {Yoshioka}, \citenamefont {Sagawa},\ and\ \citenamefont {Lukin}}]{yada2024experimentally}%
  \BibitemOpen
  \bibfield  {author} {\bibinfo {author} {\bibfnamefont {T.}~\bibnamefont {Yada}}, \bibinfo {author} {\bibfnamefont {P.-J.}\ \bibnamefont {Stas}}, \bibinfo {author} {\bibfnamefont {A.}~\bibnamefont {Suleymanzade}}, \bibinfo {author} {\bibfnamefont {E.~N.}\ \bibnamefont {Knall}}, \bibinfo {author} {\bibfnamefont {N.}~\bibnamefont {Yoshioka}}, \bibinfo {author} {\bibfnamefont {T.}~\bibnamefont {Sagawa}},\ and\ \bibinfo {author} {\bibfnamefont {M.~D.}\ \bibnamefont {Lukin}},\ }\bibfield  {title} {\bibinfo {title} {Experimentally probing entropy reduction via iterative quantum information transfer},\ }\href {https://doi.org/10.1103/5lp2-9sps} {\bibfield  {journal} {\bibinfo  {journal} {Phys. Rev. X}\ }\textbf {\bibinfo {volume} {15}},\ \bibinfo {pages} {031054} (\bibinfo {year} {2025})}\BibitemShut {NoStop}%
\bibitem [{\citenamefont {Prech}\ \emph {et~al.}(2025)\citenamefont {Prech}, \citenamefont {Aschwanden},\ and\ \citenamefont {Potts}}]{prech2025quantum}%
  \BibitemOpen
  \bibfield  {author} {\bibinfo {author} {\bibfnamefont {K.}~\bibnamefont {Prech}}, \bibinfo {author} {\bibfnamefont {J.}~\bibnamefont {Aschwanden}},\ and\ \bibinfo {author} {\bibfnamefont {P.~P.}\ \bibnamefont {Potts}},\ }\bibfield  {title} {\bibinfo {title} {Quantum thermodynamics of continuous feedback control},\ }\href@noop {} {\bibfield  {journal} {\bibinfo  {journal} {arXiv:2505.16615}\ } (\bibinfo {year} {2025})}\BibitemShut {NoStop}%
\bibitem [{\citenamefont {Barato}\ and\ \citenamefont {Seifert}(2015)}]{Barato_and_Seifert_2015_PhysRevLett.114.158101}%
  \BibitemOpen
  \bibfield  {author} {\bibinfo {author} {\bibfnamefont {A.~C.}\ \bibnamefont {Barato}}\ and\ \bibinfo {author} {\bibfnamefont {U.}~\bibnamefont {Seifert}},\ }\bibfield  {title} {\bibinfo {title} {Thermodynamic uncertainty relation for biomolecular processes},\ }\href {https://doi.org/10.1103/PhysRevLett.114.158101} {\bibfield  {journal} {\bibinfo  {journal} {Phys. Rev. Lett.}\ }\textbf {\bibinfo {volume} {114}},\ \bibinfo {pages} {158101} (\bibinfo {year} {2015})}\BibitemShut {NoStop}%
\bibitem [{\citenamefont {Gingrich}\ \emph {et~al.}(2016)\citenamefont {Gingrich}, \citenamefont {Horowitz}, \citenamefont {Perunov},\ and\ \citenamefont {England}}]{Gingrich_2016_PhysRevLett.116.120601}%
  \BibitemOpen
  \bibfield  {author} {\bibinfo {author} {\bibfnamefont {T.~R.}\ \bibnamefont {Gingrich}}, \bibinfo {author} {\bibfnamefont {J.~M.}\ \bibnamefont {Horowitz}}, \bibinfo {author} {\bibfnamefont {N.}~\bibnamefont {Perunov}},\ and\ \bibinfo {author} {\bibfnamefont {J.~L.}\ \bibnamefont {England}},\ }\bibfield  {title} {\bibinfo {title} {Dissipation bounds all steady-state current fluctuations},\ }\href {https://doi.org/10.1103/PhysRevLett.116.120601} {\bibfield  {journal} {\bibinfo  {journal} {Phys. Rev. Lett.}\ }\textbf {\bibinfo {volume} {116}},\ \bibinfo {pages} {120601} (\bibinfo {year} {2016})}\BibitemShut {NoStop}%
\bibitem [{\citenamefont {Horowitz}\ and\ \citenamefont {Gingrich}(2017)}]{Horowitz_and_Gingrich_2017_PhysRevE.96.020103}%
  \BibitemOpen
  \bibfield  {author} {\bibinfo {author} {\bibfnamefont {J.~M.}\ \bibnamefont {Horowitz}}\ and\ \bibinfo {author} {\bibfnamefont {T.~R.}\ \bibnamefont {Gingrich}},\ }\bibfield  {title} {\bibinfo {title} {Proof of the finite-time thermodynamic uncertainty relation for steady-state currents},\ }\href {https://doi.org/10.1103/PhysRevE.96.020103} {\bibfield  {journal} {\bibinfo  {journal} {Phys. Rev. E}\ }\textbf {\bibinfo {volume} {96}},\ \bibinfo {pages} {020103} (\bibinfo {year} {2017})}\BibitemShut {NoStop}%
\bibitem [{\citenamefont {Dechant}\ and\ \citenamefont {Sasa}(2018)}]{Dechant_2018}%
  \BibitemOpen
  \bibfield  {author} {\bibinfo {author} {\bibfnamefont {A.}~\bibnamefont {Dechant}}\ and\ \bibinfo {author} {\bibfnamefont {S.-i.}\ \bibnamefont {Sasa}},\ }\bibfield  {title} {\bibinfo {title} {Current fluctuations and transport efficiency for general langevin systems},\ }\href {https://doi.org/10.1088/1742-5468/aac91a} {\bibfield  {journal} {\bibinfo  {journal} {Journal of Statistical Mechanics: Theory and Experiment}\ }\textbf {\bibinfo {volume} {2018}},\ \bibinfo {pages} {063209} (\bibinfo {year} {2018})}\BibitemShut {NoStop}%
\bibitem [{\citenamefont {Dechant}(2018)}]{Dechant_2018_2}%
  \BibitemOpen
  \bibfield  {author} {\bibinfo {author} {\bibfnamefont {A.}~\bibnamefont {Dechant}},\ }\bibfield  {title} {\bibinfo {title} {Multidimensional thermodynamic uncertainty relations},\ }\href {https://doi.org/10.1088/1751-8121/aaf3ff} {\bibfield  {journal} {\bibinfo  {journal} {Journal of Physics A: Mathematical and Theoretical}\ }\textbf {\bibinfo {volume} {52}},\ \bibinfo {pages} {035001} (\bibinfo {year} {2018})}\BibitemShut {NoStop}%
\bibitem [{\citenamefont {Van~Vu}\ and\ \citenamefont {Hasegawa}(2019)}]{Vu_and_Hasegawa_2019_PhysRevE.100.032130}%
  \BibitemOpen
  \bibfield  {author} {\bibinfo {author} {\bibfnamefont {T.}~\bibnamefont {Van~Vu}}\ and\ \bibinfo {author} {\bibfnamefont {Y.}~\bibnamefont {Hasegawa}},\ }\bibfield  {title} {\bibinfo {title} {Uncertainty relations for underdamped langevin dynamics},\ }\href {https://doi.org/10.1103/PhysRevE.100.032130} {\bibfield  {journal} {\bibinfo  {journal} {Phys. Rev. E}\ }\textbf {\bibinfo {volume} {100}},\ \bibinfo {pages} {032130} (\bibinfo {year} {2019})}\BibitemShut {NoStop}%
\bibitem [{\citenamefont {Lee}\ \emph {et~al.}(2019)\citenamefont {Lee}, \citenamefont {Park},\ and\ \citenamefont {Park}}]{Lee_2019_PhysRevE.100.062132}%
  \BibitemOpen
  \bibfield  {author} {\bibinfo {author} {\bibfnamefont {J.~S.}\ \bibnamefont {Lee}}, \bibinfo {author} {\bibfnamefont {J.-M.}\ \bibnamefont {Park}},\ and\ \bibinfo {author} {\bibfnamefont {H.}~\bibnamefont {Park}},\ }\bibfield  {title} {\bibinfo {title} {Thermodynamic uncertainty relation for underdamped langevin systems driven by a velocity-dependent force},\ }\href {https://doi.org/10.1103/PhysRevE.100.062132} {\bibfield  {journal} {\bibinfo  {journal} {Phys. Rev. E}\ }\textbf {\bibinfo {volume} {100}},\ \bibinfo {pages} {062132} (\bibinfo {year} {2019})}\BibitemShut {NoStop}%
\bibitem [{\citenamefont {Shiraishi}\ \emph {et~al.}(2016)\citenamefont {Shiraishi}, \citenamefont {Saito},\ and\ \citenamefont {Tasaki}}]{Shiraishi_2016_PhysRevLett.117.190601}%
  \BibitemOpen
  \bibfield  {author} {\bibinfo {author} {\bibfnamefont {N.}~\bibnamefont {Shiraishi}}, \bibinfo {author} {\bibfnamefont {K.}~\bibnamefont {Saito}},\ and\ \bibinfo {author} {\bibfnamefont {H.}~\bibnamefont {Tasaki}},\ }\bibfield  {title} {\bibinfo {title} {Universal trade-off relation between power and efficiency for heat engines},\ }\href {https://doi.org/10.1103/PhysRevLett.117.190601} {\bibfield  {journal} {\bibinfo  {journal} {Phys. Rev. Lett.}\ }\textbf {\bibinfo {volume} {117}},\ \bibinfo {pages} {190601} (\bibinfo {year} {2016})}\BibitemShut {NoStop}%
\bibitem [{\citenamefont {Pietzonka}\ and\ \citenamefont {Seifert}(2018)}]{Pietzonka_and_Seifert_2018_PhysRevLett.120.190602}%
  \BibitemOpen
  \bibfield  {author} {\bibinfo {author} {\bibfnamefont {P.}~\bibnamefont {Pietzonka}}\ and\ \bibinfo {author} {\bibfnamefont {U.}~\bibnamefont {Seifert}},\ }\bibfield  {title} {\bibinfo {title} {Universal trade-off between power, efficiency, and constancy in steady-state heat engines},\ }\href {https://doi.org/10.1103/PhysRevLett.120.190602} {\bibfield  {journal} {\bibinfo  {journal} {Phys. Rev. Lett.}\ }\textbf {\bibinfo {volume} {120}},\ \bibinfo {pages} {190602} (\bibinfo {year} {2018})}\BibitemShut {NoStop}%
\bibitem [{\citenamefont {Tajima}\ and\ \citenamefont {Funo}(2021)}]{PhysRevLett.127.190604}%
  \BibitemOpen
  \bibfield  {author} {\bibinfo {author} {\bibfnamefont {H.}~\bibnamefont {Tajima}}\ and\ \bibinfo {author} {\bibfnamefont {K.}~\bibnamefont {Funo}},\ }\bibfield  {title} {\bibinfo {title} {Superconducting-like heat current: Effective cancellation of current-dissipation trade-off by quantum coherence},\ }\href {https://doi.org/10.1103/PhysRevLett.127.190604} {\bibfield  {journal} {\bibinfo  {journal} {Phys. Rev. Lett.}\ }\textbf {\bibinfo {volume} {127}},\ \bibinfo {pages} {190604} (\bibinfo {year} {2021})}\BibitemShut {NoStop}%
\bibitem [{\citenamefont {Potts}\ and\ \citenamefont {Samuelsson}(2019)}]{Potts_2019}%
  \BibitemOpen
  \bibfield  {author} {\bibinfo {author} {\bibfnamefont {P.~P.}\ \bibnamefont {Potts}}\ and\ \bibinfo {author} {\bibfnamefont {P.}~\bibnamefont {Samuelsson}},\ }\bibfield  {title} {\bibinfo {title} {Thermodynamic uncertainty relations including measurement and feedback},\ }\href {https://doi.org/10.1103/PhysRevE.100.052137} {\bibfield  {journal} {\bibinfo  {journal} {Phys. Rev. E}\ }\textbf {\bibinfo {volume} {100}},\ \bibinfo {pages} {052137} (\bibinfo {year} {2019})}\BibitemShut {NoStop}%
\bibitem [{\citenamefont {Van~Vu}\ and\ \citenamefont {Hasegawa}(2020)}]{Van_Vu_2020}%
  \BibitemOpen
  \bibfield  {author} {\bibinfo {author} {\bibfnamefont {T.}~\bibnamefont {Van~Vu}}\ and\ \bibinfo {author} {\bibfnamefont {Y.}~\bibnamefont {Hasegawa}},\ }\bibfield  {title} {\bibinfo {title} {Uncertainty relation under information measurement and feedback control},\ }\href {https://doi.org/10.1088/1751-8121/ab64a4} {\bibfield  {journal} {\bibinfo  {journal} {Journal of Physics A: Mathematical and Theoretical}\ }\textbf {\bibinfo {volume} {53}},\ \bibinfo {pages} {075001} (\bibinfo {year} {2020})}\BibitemShut {NoStop}%
\bibitem [{\citenamefont {Liu}\ \emph {et~al.}(2020)\citenamefont {Liu}, \citenamefont {Gong},\ and\ \citenamefont {Ueda}}]{Liu_2020}%
  \BibitemOpen
  \bibfield  {author} {\bibinfo {author} {\bibfnamefont {K.}~\bibnamefont {Liu}}, \bibinfo {author} {\bibfnamefont {Z.}~\bibnamefont {Gong}},\ and\ \bibinfo {author} {\bibfnamefont {M.}~\bibnamefont {Ueda}},\ }\bibfield  {title} {\bibinfo {title} {Thermodynamic uncertainty relation for arbitrary initial states},\ }\href {https://doi.org/10.1103/PhysRevLett.125.140602} {\bibfield  {journal} {\bibinfo  {journal} {Phys. Rev. Lett.}\ }\textbf {\bibinfo {volume} {125}},\ \bibinfo {pages} {140602} (\bibinfo {year} {2020})}\BibitemShut {NoStop}%
\bibitem [{\citenamefont {Otsubo}\ \emph {et~al.}(2020)\citenamefont {Otsubo}, \citenamefont {Ito}, \citenamefont {Dechant},\ and\ \citenamefont {Sagawa}}]{Otsubo_PhysRevE.101.062106}%
  \BibitemOpen
  \bibfield  {author} {\bibinfo {author} {\bibfnamefont {S.}~\bibnamefont {Otsubo}}, \bibinfo {author} {\bibfnamefont {S.}~\bibnamefont {Ito}}, \bibinfo {author} {\bibfnamefont {A.}~\bibnamefont {Dechant}},\ and\ \bibinfo {author} {\bibfnamefont {T.}~\bibnamefont {Sagawa}},\ }\bibfield  {title} {\bibinfo {title} {Estimating entropy production by machine learning of short-time fluctuating currents},\ }\href {https://doi.org/10.1103/PhysRevE.101.062106} {\bibfield  {journal} {\bibinfo  {journal} {Phys. Rev. E}\ }\textbf {\bibinfo {volume} {101}},\ \bibinfo {pages} {062106} (\bibinfo {year} {2020})}\BibitemShut {NoStop}%
\bibitem [{\citenamefont {Tanogami}\ \emph {et~al.}(2023)\citenamefont {Tanogami}, \citenamefont {Van~Vu},\ and\ \citenamefont {Saito}}]{Tanogami_PhysRevResearch.5.043280}%
  \BibitemOpen
  \bibfield  {author} {\bibinfo {author} {\bibfnamefont {T.}~\bibnamefont {Tanogami}}, \bibinfo {author} {\bibfnamefont {T.}~\bibnamefont {Van~Vu}},\ and\ \bibinfo {author} {\bibfnamefont {K.}~\bibnamefont {Saito}},\ }\bibfield  {title} {\bibinfo {title} {Universal bounds on the performance of information-thermodynamic engine},\ }\href {https://doi.org/10.1103/PhysRevResearch.5.043280} {\bibfield  {journal} {\bibinfo  {journal} {Phys. Rev. Res.}\ }\textbf {\bibinfo {volume} {5}},\ \bibinfo {pages} {043280} (\bibinfo {year} {2023})}\BibitemShut {NoStop}%
\bibitem [{\citenamefont {Rossi}\ \emph {et~al.}(2018)\citenamefont {Rossi}, \citenamefont {Mason}, \citenamefont {Chen}, \citenamefont {Tsaturyan},\ and\ \citenamefont {Schliesser}}]{Rossi_2018}%
  \BibitemOpen
  \bibfield  {author} {\bibinfo {author} {\bibfnamefont {M.}~\bibnamefont {Rossi}}, \bibinfo {author} {\bibfnamefont {D.}~\bibnamefont {Mason}}, \bibinfo {author} {\bibfnamefont {J.}~\bibnamefont {Chen}}, \bibinfo {author} {\bibfnamefont {Y.}~\bibnamefont {Tsaturyan}},\ and\ \bibinfo {author} {\bibfnamefont {A.}~\bibnamefont {Schliesser}},\ }\bibfield  {title} {\bibinfo {title} {Measurement-based quantum control of mechanical motion},\ }\href {https://doi.org/10.1038/s41586-018-0643-8} {\bibfield  {journal} {\bibinfo  {journal} {Nature}\ }\textbf {\bibinfo {volume} {563}},\ \bibinfo {pages} {53} (\bibinfo {year} {2018})}\BibitemShut {NoStop}%
\bibitem [{\citenamefont {Gieseler}\ \emph {et~al.}(2012)\citenamefont {Gieseler}, \citenamefont {Deutsch}, \citenamefont {Quidant},\ and\ \citenamefont {Novotny}}]{Ginseler_2012_PhysRevLett.109.103603}%
  \BibitemOpen
  \bibfield  {author} {\bibinfo {author} {\bibfnamefont {J.}~\bibnamefont {Gieseler}}, \bibinfo {author} {\bibfnamefont {B.}~\bibnamefont {Deutsch}}, \bibinfo {author} {\bibfnamefont {R.}~\bibnamefont {Quidant}},\ and\ \bibinfo {author} {\bibfnamefont {L.}~\bibnamefont {Novotny}},\ }\bibfield  {title} {\bibinfo {title} {Subkelvin parametric feedback cooling of a laser-trapped nanoparticle},\ }\href {https://doi.org/10.1103/PhysRevLett.109.103603} {\bibfield  {journal} {\bibinfo  {journal} {Phys. Rev. Lett.}\ }\textbf {\bibinfo {volume} {109}},\ \bibinfo {pages} {103603} (\bibinfo {year} {2012})}\BibitemShut {NoStop}%
\bibitem [{\citenamefont {Magrini}\ \emph {et~al.}(2021)\citenamefont {Magrini}, \citenamefont {Rosenzweig}, \citenamefont {Bach}, \citenamefont {Deutschmann-Olek}, \citenamefont {Hofer}, \citenamefont {Hong}, \citenamefont {Kiesel}, \citenamefont {Kugi},\ and\ \citenamefont {Aspelmeyer}}]{Magrini_2021}%
  \BibitemOpen
  \bibfield  {author} {\bibinfo {author} {\bibfnamefont {L.}~\bibnamefont {Magrini}}, \bibinfo {author} {\bibfnamefont {P.}~\bibnamefont {Rosenzweig}}, \bibinfo {author} {\bibfnamefont {C.}~\bibnamefont {Bach}}, \bibinfo {author} {\bibfnamefont {A.}~\bibnamefont {Deutschmann-Olek}}, \bibinfo {author} {\bibfnamefont {S.~G.}\ \bibnamefont {Hofer}}, \bibinfo {author} {\bibfnamefont {S.}~\bibnamefont {Hong}}, \bibinfo {author} {\bibfnamefont {N.}~\bibnamefont {Kiesel}}, \bibinfo {author} {\bibfnamefont {A.}~\bibnamefont {Kugi}},\ and\ \bibinfo {author} {\bibfnamefont {M.}~\bibnamefont {Aspelmeyer}},\ }\bibfield  {title} {\bibinfo {title} {Real-time optimal quantum control of mechanical motion at room temperature},\ }\href {https://doi.org/10.1038/s41586-021-03602-3} {\bibfield  {journal} {\bibinfo  {journal} {Nature}\ }\textbf {\bibinfo {volume} {595}},\ \bibinfo {pages} {373} (\bibinfo {year} {2021})}\BibitemShut {NoStop}%
\bibitem [{\citenamefont {Gonzalez-Ballestero}\ \emph {et~al.}(2021)\citenamefont {Gonzalez-Ballestero}, \citenamefont {Aspelmeyer}, \citenamefont {Novotny}, \citenamefont {Quidant},\ and\ \citenamefont {Romero-Isart}}]{Gonzalez_Ballestero_2021}%
  \BibitemOpen
  \bibfield  {author} {\bibinfo {author} {\bibfnamefont {C.}~\bibnamefont {Gonzalez-Ballestero}}, \bibinfo {author} {\bibfnamefont {M.}~\bibnamefont {Aspelmeyer}}, \bibinfo {author} {\bibfnamefont {L.}~\bibnamefont {Novotny}}, \bibinfo {author} {\bibfnamefont {R.}~\bibnamefont {Quidant}},\ and\ \bibinfo {author} {\bibfnamefont {O.}~\bibnamefont {Romero-Isart}},\ }\bibfield  {title} {\bibinfo {title} {Levitodynamics: Levitation and control of microscopic objects in vacuum},\ }\href {https://doi.org/10.1126/science.abg3027} {\bibfield  {journal} {\bibinfo  {journal} {Science}\ }\textbf {\bibinfo {volume} {374}},\ \bibinfo {pages} {eabg3027} (\bibinfo {year} {2021})}\BibitemShut {NoStop}%
\bibitem [{\citenamefont {Kamba}\ \emph {et~al.}(2023)\citenamefont {Kamba}, \citenamefont {Shimizu},\ and\ \citenamefont {Aikawa}}]{Kamba_shimizu_aikawa_2023}%
  \BibitemOpen
  \bibfield  {author} {\bibinfo {author} {\bibfnamefont {M.}~\bibnamefont {Kamba}}, \bibinfo {author} {\bibfnamefont {R.}~\bibnamefont {Shimizu}},\ and\ \bibinfo {author} {\bibfnamefont {K.}~\bibnamefont {Aikawa}},\ }\bibfield  {title} {\bibinfo {title} {Nanoscale feedback control of six degrees of freedom of a near-sphere},\ }\href {https://doi.org/10.1038/s41467-023-43745-7} {\bibfield  {journal} {\bibinfo  {journal} {Nature Communications}\ }\textbf {\bibinfo {volume} {14}},\ \bibinfo {pages} {7943} (\bibinfo {year} {2023})}\BibitemShut {NoStop}%
\bibitem [{\citenamefont {Kamba}\ \emph {et~al.}(2025)\citenamefont {Kamba}, \citenamefont {Hara},\ and\ \citenamefont {Aikawa}}]{kamba2025quantum}%
  \BibitemOpen
  \bibfield  {author} {\bibinfo {author} {\bibfnamefont {M.}~\bibnamefont {Kamba}}, \bibinfo {author} {\bibfnamefont {N.}~\bibnamefont {Hara}},\ and\ \bibinfo {author} {\bibfnamefont {K.}~\bibnamefont {Aikawa}},\ }\bibfield  {title} {\bibinfo {title} {Quantum squeezing of a levitated nanomechanical oscillator},\ }\href {https://doi.org/10.1126/science.ady4652} {\bibfield  {journal} {\bibinfo  {journal} {Science}\ }\textbf {\bibinfo {volume} {389}},\ \bibinfo {pages} {1225} (\bibinfo {year} {2025})}\BibitemShut {NoStop}%
\bibitem [{\citenamefont {Horowitz}\ and\ \citenamefont {Sandberg}(2014)}]{Horowitz_and_Sandberg_2014}%
  \BibitemOpen
  \bibfield  {author} {\bibinfo {author} {\bibfnamefont {J.~M.}\ \bibnamefont {Horowitz}}\ and\ \bibinfo {author} {\bibfnamefont {H.}~\bibnamefont {Sandberg}},\ }\bibfield  {title} {\bibinfo {title} {Second-law-like inequalities with information and their interpretations},\ }\href {https://doi.org/10.1088/1367-2630/16/12/125007} {\bibfield  {journal} {\bibinfo  {journal} {New Journal of Physics}\ }\textbf {\bibinfo {volume} {16}},\ \bibinfo {pages} {125007} (\bibinfo {year} {2014})}\BibitemShut {NoStop}%
\bibitem [{\citenamefont {Sandberg}\ \emph {et~al.}(2014)\citenamefont {Sandberg}, \citenamefont {Delvenne}, \citenamefont {Newton},\ and\ \citenamefont {Mitter}}]{Sandberg_2014_PhysRevE.90.042119}%
  \BibitemOpen
  \bibfield  {author} {\bibinfo {author} {\bibfnamefont {H.}~\bibnamefont {Sandberg}}, \bibinfo {author} {\bibfnamefont {J.-C.}\ \bibnamefont {Delvenne}}, \bibinfo {author} {\bibfnamefont {N.~J.}\ \bibnamefont {Newton}},\ and\ \bibinfo {author} {\bibfnamefont {S.~K.}\ \bibnamefont {Mitter}},\ }\bibfield  {title} {\bibinfo {title} {Maximum work extraction and implementation costs for nonequilibrium maxwell's demons},\ }\href {https://doi.org/10.1103/PhysRevE.90.042119} {\bibfield  {journal} {\bibinfo  {journal} {Phys. Rev. E}\ }\textbf {\bibinfo {volume} {90}},\ \bibinfo {pages} {042119} (\bibinfo {year} {2014})}\BibitemShut {NoStop}%
\bibitem [{\citenamefont {Schreiber}(2000)}]{PhysRevLett.85.461}%
  \BibitemOpen
  \bibfield  {author} {\bibinfo {author} {\bibfnamefont {T.}~\bibnamefont {Schreiber}},\ }\bibfield  {title} {\bibinfo {title} {Measuring information transfer},\ }\href {https://doi.org/10.1103/PhysRevLett.85.461} {\bibfield  {journal} {\bibinfo  {journal} {Phys. Rev. Lett.}\ }\textbf {\bibinfo {volume} {85}},\ \bibinfo {pages} {461} (\bibinfo {year} {2000})}\BibitemShut {NoStop}%
\bibitem [{\citenamefont {Hartich}\ \emph {et~al.}(2014)\citenamefont {Hartich}, \citenamefont {Barato},\ and\ \citenamefont {Seifert}}]{Hartich_2014}%
  \BibitemOpen
  \bibfield  {author} {\bibinfo {author} {\bibfnamefont {D.}~\bibnamefont {Hartich}}, \bibinfo {author} {\bibfnamefont {A.~C.}\ \bibnamefont {Barato}},\ and\ \bibinfo {author} {\bibfnamefont {U.}~\bibnamefont {Seifert}},\ }\bibfield  {title} {\bibinfo {title} {Stochastic thermodynamics of bipartite systems: transfer entropy inequalities and a maxwell’s demon interpretation},\ }\href {https://doi.org/10.1088/1742-5468/2014/02/P02016} {\bibfield  {journal} {\bibinfo  {journal} {Journal of Statistical Mechanics: Theory and Experiment}\ }\textbf {\bibinfo {volume} {2014}},\ \bibinfo {pages} {P02016} (\bibinfo {year} {2014})}\BibitemShut {NoStop}%
\bibitem [{\citenamefont {Kumasaki}\ \emph {et~al.}(2025)\citenamefont {Kumasaki}, \citenamefont {Yada}, \citenamefont {Funo},\ and\ \citenamefont {Sagawa}}]{kumasaki2025}%
  \BibitemOpen
  \bibfield  {author} {\bibinfo {author} {\bibfnamefont {K.}~\bibnamefont {Kumasaki}}, \bibinfo {author} {\bibfnamefont {T.}~\bibnamefont {Yada}}, \bibinfo {author} {\bibfnamefont {K.}~\bibnamefont {Funo}},\ and\ \bibinfo {author} {\bibfnamefont {T.}~\bibnamefont {Sagawa}},\ }\bibfield  {title} {\bibinfo {title} {Thermodynamic approach to quantum cooling limit of continuous gaussian feedback},\ }\href {https://doi.org/10.1103/5cz5-n6jt} {\bibfield  {journal} {\bibinfo  {journal} {Phys. Rev. Res.}\ }\textbf {\bibinfo {volume} {7}},\ \bibinfo {pages} {043147} (\bibinfo {year} {2025})}\BibitemShut {NoStop}%
\bibitem [{\citenamefont {Dechant}\ \emph {et~al.}(2022{\natexlab{a}})\citenamefont {Dechant}, \citenamefont {Sasa},\ and\ \citenamefont {Ito}}]{Dechant_PhysRevE.106.024125}%
  \BibitemOpen
  \bibfield  {author} {\bibinfo {author} {\bibfnamefont {A.}~\bibnamefont {Dechant}}, \bibinfo {author} {\bibfnamefont {S.-i.}\ \bibnamefont {Sasa}},\ and\ \bibinfo {author} {\bibfnamefont {S.}~\bibnamefont {Ito}},\ }\bibfield  {title} {\bibinfo {title} {Geometric decomposition of entropy production into excess, housekeeping, and coupling parts},\ }\href {https://doi.org/10.1103/PhysRevE.106.024125} {\bibfield  {journal} {\bibinfo  {journal} {Phys. Rev. E}\ }\textbf {\bibinfo {volume} {106}},\ \bibinfo {pages} {024125} (\bibinfo {year} {2022}{\natexlab{a}})}\BibitemShut {NoStop}%
\bibitem [{\citenamefont {Dechant}\ \emph {et~al.}(2022{\natexlab{b}})\citenamefont {Dechant}, \citenamefont {Sasa},\ and\ \citenamefont {Ito}}]{Dechant_2022_2}%
  \BibitemOpen
  \bibfield  {author} {\bibinfo {author} {\bibfnamefont {A.}~\bibnamefont {Dechant}}, \bibinfo {author} {\bibfnamefont {S.-i.}\ \bibnamefont {Sasa}},\ and\ \bibinfo {author} {\bibfnamefont {S.}~\bibnamefont {Ito}},\ }\bibfield  {title} {\bibinfo {title} {Geometric decomposition of entropy production in out-of-equilibrium systems},\ }\href {https://doi.org/10.1103/PhysRevResearch.4.L012034} {\bibfield  {journal} {\bibinfo  {journal} {Phys. Rev. Res.}\ }\textbf {\bibinfo {volume} {4}},\ \bibinfo {pages} {L012034} (\bibinfo {year} {2022}{\natexlab{b}})}\BibitemShut {NoStop}%
\bibitem [{\citenamefont {Kamijima}\ \emph {et~al.}(2023)\citenamefont {Kamijima}, \citenamefont {Ito}, \citenamefont {Dechant},\ and\ \citenamefont {Sagawa}}]{Kamijima_2023_PhysRevE.107.L052101}%
  \BibitemOpen
  \bibfield  {author} {\bibinfo {author} {\bibfnamefont {T.}~\bibnamefont {Kamijima}}, \bibinfo {author} {\bibfnamefont {S.}~\bibnamefont {Ito}}, \bibinfo {author} {\bibfnamefont {A.}~\bibnamefont {Dechant}},\ and\ \bibinfo {author} {\bibfnamefont {T.}~\bibnamefont {Sagawa}},\ }\bibfield  {title} {\bibinfo {title} {Thermodynamic uncertainty relations for steady-state thermodynamics},\ }\href {https://doi.org/10.1103/PhysRevE.107.L052101} {\bibfield  {journal} {\bibinfo  {journal} {Phys. Rev. E}\ }\textbf {\bibinfo {volume} {107}},\ \bibinfo {pages} {L052101} (\bibinfo {year} {2023})}\BibitemShut {NoStop}%
\bibitem [{\citenamefont {Munakata}\ and\ \citenamefont {Rosinberg}(2013)}]{Munakata_2013}%
  \BibitemOpen
  \bibfield  {author} {\bibinfo {author} {\bibfnamefont {T.}~\bibnamefont {Munakata}}\ and\ \bibinfo {author} {\bibfnamefont {M.~L.}\ \bibnamefont {Rosinberg}},\ }\bibfield  {title} {\bibinfo {title} {Feedback cooling, measurement errors, and entropy production},\ }\href {https://doi.org/10.1088/1742-5468/2013/06/P06014} {\bibfield  {journal} {\bibinfo  {journal} {Journal of Statistical Mechanics: Theory and Experiment}\ }\textbf {\bibinfo {volume} {2013}},\ \bibinfo {pages} {P06014} (\bibinfo {year} {2013})}\BibitemShut {NoStop}%
\bibitem [{\citenamefont {Risken}(1996)}]{Risken1996}%
  \BibitemOpen
  \bibfield  {author} {\bibinfo {author} {\bibfnamefont {H.}~\bibnamefont {Risken}},\ }\bibinfo {title} {Fokker-planck equation},\ in\ \href {https://doi.org/10.1007/978-3-642-61544-3_4} {\emph {\bibinfo {booktitle} {The Fokker-Planck Equation: Methods of Solution and Applications}}}\ (\bibinfo  {publisher} {Springer Berlin Heidelberg},\ \bibinfo {address} {Berlin, Heidelberg},\ \bibinfo {year} {1996})\ pp.\ \bibinfo {pages} {63--95}\BibitemShut {NoStop}%
\bibitem [{\citenamefont {Spinney}\ and\ \citenamefont {Ford}(2012)}]{Spinney2012}%
  \BibitemOpen
  \bibfield  {author} {\bibinfo {author} {\bibfnamefont {R.~E.}\ \bibnamefont {Spinney}}\ and\ \bibinfo {author} {\bibfnamefont {I.~J.}\ \bibnamefont {Ford}},\ }\bibfield  {title} {\bibinfo {title} {Entropy production in full phase space for continuous stochastic dynamics},\ }\href {https://doi.org/10.1103/PhysRevE.85.051113} {\bibfield  {journal} {\bibinfo  {journal} {Phys. Rev. E}\ }\textbf {\bibinfo {volume} {85}},\ \bibinfo {pages} {051113} (\bibinfo {year} {2012})}\BibitemShut {NoStop}%
\bibitem [{\citenamefont {Rosinberg}\ \emph {et~al.}(2015)\citenamefont {Rosinberg}, \citenamefont {Munakata},\ and\ \citenamefont {Tarjus}}]{Rosinberg2015}%
  \BibitemOpen
  \bibfield  {author} {\bibinfo {author} {\bibfnamefont {M.~L.}\ \bibnamefont {Rosinberg}}, \bibinfo {author} {\bibfnamefont {T.}~\bibnamefont {Munakata}},\ and\ \bibinfo {author} {\bibfnamefont {G.}~\bibnamefont {Tarjus}},\ }\bibfield  {title} {\bibinfo {title} {Stochastic thermodynamics of langevin systems under time-delayed feedback control: Second-law-like inequalities},\ }\href {https://doi.org/10.1103/PhysRevE.91.042114} {\bibfield  {journal} {\bibinfo  {journal} {Phys. Rev. E}\ }\textbf {\bibinfo {volume} {91}},\ \bibinfo {pages} {042114} (\bibinfo {year} {2015})}\BibitemShut {NoStop}%
\bibitem [{\citenamefont {Rosinberg}\ and\ \citenamefont {Horowitz}(2016)}]{Rosinberg_2016}%
  \BibitemOpen
  \bibfield  {author} {\bibinfo {author} {\bibfnamefont {M.~L.}\ \bibnamefont {Rosinberg}}\ and\ \bibinfo {author} {\bibfnamefont {J.~M.}\ \bibnamefont {Horowitz}},\ }\bibfield  {title} {\bibinfo {title} {Continuous information flow fluctuations},\ }\href {https://doi.org/10.1209/0295-5075/116/10007} {\bibfield  {journal} {\bibinfo  {journal} {Europhysics Letters}\ }\textbf {\bibinfo {volume} {116}},\ \bibinfo {pages} {10007} (\bibinfo {year} {2016})}\BibitemShut {NoStop}%
\bibitem [{\citenamefont {Hatano}\ and\ \citenamefont {Sasa}(2001)}]{Hatano_and_Sasa_2001_PhysRevLett.86.3463}%
  \BibitemOpen
  \bibfield  {author} {\bibinfo {author} {\bibfnamefont {T.}~\bibnamefont {Hatano}}\ and\ \bibinfo {author} {\bibfnamefont {S.-i.}\ \bibnamefont {Sasa}},\ }\bibfield  {title} {\bibinfo {title} {Steady-state thermodynamics of langevin systems},\ }\href {https://doi.org/10.1103/PhysRevLett.86.3463} {\bibfield  {journal} {\bibinfo  {journal} {Phys. Rev. Lett.}\ }\textbf {\bibinfo {volume} {86}},\ \bibinfo {pages} {3463} (\bibinfo {year} {2001})}\BibitemShut {NoStop}%
\bibitem [{\citenamefont {Maes}\ and\ \citenamefont {Neto{\v{c}}n{\'y}}(2014)}]{Maes_and_Netocny_2014}%
  \BibitemOpen
  \bibfield  {author} {\bibinfo {author} {\bibfnamefont {C.}~\bibnamefont {Maes}}\ and\ \bibinfo {author} {\bibfnamefont {K.}~\bibnamefont {Neto{\v{c}}n{\'y}}},\ }\bibfield  {title} {\bibinfo {title} {A nonequilibrium extension of the clausius heat theorem},\ }\href {https://doi.org/10.1007/s10955-013-0822-9} {\bibfield  {journal} {\bibinfo  {journal} {Journal of Statistical Physics}\ }\textbf {\bibinfo {volume} {154}},\ \bibinfo {pages} {188} (\bibinfo {year} {2014})}\BibitemShut {NoStop}%
\bibitem [{\citenamefont {Simon}(2006)}]{simon2006optimal}%
  \BibitemOpen
  \bibfield  {author} {\bibinfo {author} {\bibfnamefont {D.}~\bibnamefont {Simon}},\ }\href@noop {} {\emph {\bibinfo {title} {Optimal state estimation: Kalman, H infinity, and nonlinear approaches}}}\ (\bibinfo  {publisher} {John Wiley \& Sons},\ \bibinfo {year} {2006})\BibitemShut {NoStop}%
\end{thebibliography}
\end{document}